\documentclass[twocolumn]{aastex63}
\usepackage{comment}
\usepackage{amsmath}
\usepackage{courier}
\usepackage{float}
\usepackage{url}
\usepackage[T1]{fontenc}
\accepted{by ApJ}

\shorttitle{Environmental dependence of gas properties in a protocluster at $z\sim 2.5$}
\shortauthors{Aoyama et al.}
\graphicspath{{./}{figures/}}

\begin{document}

\title{The environmental dependence of gas properties in dense cores of a protocluster \\at $z\sim2.5$ revealed with ALMA}

%%%%% author %%%%%
\correspondingauthor{Tadayuki Kodama}
\email{kodama@astr.tohoku.ac.jp}

\author[0000-0002-6422-4594]{Kohei Aoyama}
\affiliation{Astronomical Institute, Tohoku University, 6-3 Aramaki, Aoba, Sendai, Miyagi, 980-8578, Japan}

\author[0000-0002-2993-1576]{Tadayuki Kodama}
\affiliation{Astronomical Institute, Tohoku University, 6-3 Aramaki, Aoba, Sendai, Miyagi, 980-8578, Japan}

\author[0000-0002-3560-1346]{Tomoko L. Suzuki}
\affiliation{Astronomical Institute, Tohoku University, 6-3 Aramaki, Aoba, Sendai, Miyagi, 980-8578, Japan}
\affiliation{National Astronomical Observatory of Japan, 2-21-1 Osawa, Mitaka, Tokyo, 181-8588, Japan}
\affiliation{Kavli Institute for the Physics and Mathematics of the Universe (WPI),The University of Tokyo Institutes for Advanced Study, The University of Tokyo, Kashiwa, Chiba 277-8583, Japan}
%\affiliation{Kapteyn Astronomical Institute, University of Groningen, P.O. Box 800, 9700 AV Groningen, The Netherlands}

\author[0000-0001-9728-8909]{Ken-ichi Tadaki}
\affiliation{National Astronomical Observatory of Japan, 2-21-1 Osawa, Mitaka, Tokyo, 181-8588, Japan}

\author[0000-0003-4442-2750]{Rhythm Shimakawa}
\affiliation{National Astronomical Observatory of Japan, 2-21-1 Osawa, Mitaka, Tokyo, 181-8588, Japan}

\author[0000-0002-9321-7406]{Masao Hayashi}
\affiliation{National Astronomical Observatory of Japan, 2-21-1 Osawa, Mitaka, Tokyo, 181-8588, Japan}

\author[0000-0002-0479-3699]{Yusei Koyama}
\affiliation{Subaru Telescope, National Astronomical Observatory of Japan, National Institutes of Natural Sciences (NINS), 650 North A’ohoku Place, Hilo, HI 96720, USA}
\affiliation{Department of Astronomical Science, The Graduate University for Advanced Studies (SOKENDAI), 2-21-1 Osawa, Mitaka, Tokyo, 181-8588, Japan}

\author[0000-0002-5963-6850]{
Jose Manuel P\'erez-Mart\'inez}
\affiliation{Astronomical Institute, Tohoku University, 6-3 Aramaki, Aoba, Sendai, Miyagi, 980-8578, Japan}

%%%%%%%%%%%%%%%%%%%%%%%%%%%%%%%%%

\begin{abstract}

In a protocluster USS1558-003 at $z=2.53$, galaxies in the dense cores show 
systematically elevated star-forming activities than those in less dense regions.  
To understand its origin, we look into the gas properties of the galaxies in the dense cores 
by conducting deep 1.1~mm observations with Atacama Large Millimeter/submillimeter Array (ALMA).
We detect interstellar dust continuum emissions from 12 member galaxies and estimate their molecular gas masses.
Comparing these gas masses with our previous measurements from CO(3--2) line, we infer that the latter might be overestimated. 
We find that the gas to stellar mass ratios of the galaxies
in the dense cores tend to be higher (at $M_{*}\sim 10^{10}M_{\odot}$ where we see the enhanced star-forming activities), 
suggesting that such large gas masses can sustain their high star-forming activities. 
However, if we compare the gas properties of these protocluster galaxies  
with the gas scaling relations constructed for field galaxies at similar cosmic epoch, 
we find no significant environmental difference at the same stellar mass and star formation rate. 
Although both gas mass ratios and star-forming activities are enhanced in the majority of member galaxies, they appear to follow the same scaling relation as field galaxies.  
Our results are consistent with the scenario that the cold gas is efficiently supplied to protocluster cores and to galaxies therein along surrounding filamentary structures, which leads to the high gas mass fractions and thus the elevated star-formation activities, but without changing the star formation law.

\end{abstract}  



%%%%%%%%%%%%%%%%%%%%%%%%%%%%%%%%%%%%%%%%%%%%%%%%%%

\section{Introduction} 
\label{sec:intro}

In the local Universe, it is well known that 
the properties of galaxies are strongly correlated with surrounding environments 
\citep[e.g.,][]{dressler_galaxy_1980,dressler_evolution_1997, kauffmann_environmental_2004, balogh_bimodal_2004, blanton_relationship_2005,christlein_disentangling_2005,peng_detailed_2010}.
For example, active star-forming galaxies often reside in low-density environments 
while passively evolving, older galaxies are preferentially found 
in galaxy clusters \citep[e.g.,][]{balogh_dependence_1998, poggianti_star_1999,gomez_galaxy_2003, goto_morphology-density_2003}. 
Because the majority of galaxies in clusters in the present-day Universe 
have already finished their star-formation activities, 
we need to go back in time by observing galaxies in the past 
%and see what betided in the early phase of galaxy cluster formation 
to reveal the origin of the strong environmental dependence of galaxy properties. 
At higher redshifts, active star-forming galaxies are often found 
in high-density environments, such as galaxy clusters and protoclusters 
\citep[e.g.,][]{tran_reversal_2010,fassbender_x-ray_2011,hayashi_star-bursting_2012,koyama_massive_2013,alberts_evolution_2014}. 
Star-forming activities in (proto)clusters are said to increase 
dramatically with redshift as $(1+z)^{6}$ \citep{shimakawa_identification_2014}. 
\citet{chiang_galaxy_2017} showed 
that (proto)cluster galaxies account for as much as 20\% of 
the cosmic star formation rate (SFR) density at $z=2$ 
with {\it N}-body simulations and semi-analytic models. 
(Proto)clusters at $z>2$ are ideal laboratories to see 
the early environmental effects working on member galaxies 
in the phase of the rapid growth with active star formation. 

Many environmental effects preferentially affect 
the gaseous component of cluster galaxies, subsequently star formation.
For instance, the sudden gas removal from infalling galaxies 
by ram pressure of the hot intracluster medium \citep[ICM;][]{gunn_infall_1972} 
or by galaxy harassment \citep{moore_galaxy_1996},  
and the halt of the gas supply to galaxies 
due halo gas stripping \citep[strangulation; e.g.,][]{larson_evolution_1980,kawata_strangulation_2008, peng_strangulation_2015}
are proposed as such environmental effects.
Many observations of galaxies in the local Universe reported 
a deficiency of atomic hydrogen in galaxies in denser regions 
due to ram pressure stripping \citep[e.g.,][]{vollmer_ram_2001,chung_virgo_2007, roediger_ram_2007,jaffe_budhies_2015, brown_cold_2017}. 
However, it is yet to be clarified whether molecular gas is also affected 
by ram pressure as it is more centrally concentrated and 
strongly bound to host galaxies.
Some studies showed no environmental dependence in the molecular gas fraction
\citep{kenney_effects_1989, koyama_universal_2017}.
The absence of environmental dependence may also be related to 
longer depletion timescale due to heating processes or higher turbulent pressure 
in the cluster environment \citep{mok_jcmt_2016}. 
On the other hand, some works reported 
the deficiencies of molecular gas in cluster environments 
\citep{fumagalli_molecular_2009, scott_co_2013}.
In the case of protoclusters at high redshifts, 
the situation would be different. 
Cosmological simulations suggest that 
galaxies/clusters are fed by cold gas inflow 
along the surrounding filamentary structures 
\citep[e.g.,][]{birnboim_virial_2003, keres_how_2005,dekel_galaxy_2006, ocvirk_bimodal_2008, keres_galaxies_2009,dekel_cold_2009}. 
Because galaxy clusters are located at intersections of filamentary structures, 
the gas accretion is expected to be more prominent 
in cluster environments in the distant Universe. 
Such a different mode of gas accretion in high-density environments, if any, 
could affect the gas properties of member galaxies. 

All these considerations have led us to investigate any possible environmental dependence on gas contents in galaxies at high redshifts. 
Although studies on gas contents in high-density environments at high redshifts are 
progressing, 
how the gas properties of galaxies depend on the surrounding environments at $z>2$ is inconclusive yet. 
Some studies found similar or even higher gas content in protocluster galaxies 
compared to field counterparts
\citep{lee_radio--mm_2017,zavala_gas_2019,tadaki_environmental_2019}.
In contrast, some studies reported a deficit of gas content 
in the core of matured clusters \citep{coogan_merger_2018, wang_revealing_2018}.

When we estimate the molecular gas contents in galaxies, 
CO molecular lines are commonly used as good tracers 
\citep[e.g.,][]{daddi_very_2010, genzel_study_2010, tacconi_high_2010}.
This method is well-established, but time-consuming for high redshift galaxies even with the Atacama Large Millimeter/Submillimeter Array (ALMA).
An alternative way is to observe the rest-frame far-infrared thermal emission 
from the dust associated with star-forming regions in galaxies.
Dust grains absorb rest-frame ultraviolet (UV) photons emitted from young stars
and re-emit thermal, black body-like continuum radiation in the infrared.
\citet{scoville_ism_2016} pointed out that the dust continuum emission is optically thin, 
and therefore it is a good tracer of the total amount of interstellar gas 
which hosts the dust.
Since the bulk of dust in a galaxy is cold ($T_{\rm dust} \sim 25 \rm K$), 
it can be observed at long wavelengths ($\lambda_{\rm rest} > 250 \micron$) 
where the sensitivity of ALMA is high.
Moreover, we can explore even heavily dust-obscured star-forming galaxies (DSFGs) 
which would be otherwise hidden and missed by UV-optical observations.
Therefore, we can construct a large, relatively unbiased sample of star-forming galaxies efficiently 
by deep dust continuum observations at high redshifts.

In this paper, we present the results of 
ALMA deep dust continuum imaging observations of protocluster cores at $z=2.53$ 
to study the environmental dependence of gas properties in detail.
This paper is organized as follows.
In Section \ref{sec:obs}, we explain the details of our target protocluster
and the ALMA observations. 
In Section \ref{sec:ana}, we describe our analyses 
to obtain the gas properties of our target cluster galaxies 
and the field galaxies for comparison.
Our results and discussion about the environmental dependence of gas properties 
at $z\sim2.5$ are shown in Section \ref{sec:res}.
In Section~\ref{sec:summary}, we summarize the main findings of this study. 
Throughout this paper, we assume the flat $\Lambda$CDM cosmology 
with $\Omega_{M}=0.3$, $\Omega_{\Lambda}=0.7$, and $H_{0}=70\,\rm km\,s^{-1}\,Mpc^{-1}$.
We use the \citet{chabrier_galactic_2003} initial mass function (IMF). 

%%%%%%%%%%%%%%%%%%%%%%%%%%%%%%%%%%%%%%%%%%%%%%%%%%%%%%%%%

\section{Observations}
\label{sec:obs}

\subsection{A protocluster at \texorpdfstring{$z=2.53$}{z=2.53}, USS1558-03}

A protocluster USS1558-003 (hereafter USS1558) was first discovered 
as an over-density of distant red galaxies (DRGs) 
with $J-K_{\rm s} > 2.3$ around a radio galaxy (RG), 
USS1558-03 \citep{kajisawa_protoclusters_2006}.
Narrow-band (NB) imaging observations for this protocluster 
were conducted with the Multi-Object InfraRed Camera and Spectrograph 
(MOIRCS; \citealt{ichikawa_moircs_2006,suzuki_multi-object_2008}) 
on the Subaru telescope as a part of a systematic NB imaging survey 
called MApping H-Alpha and Lines of Oxygen with Subaru 
\citep[MAHALO-Subaru;][]{kodama_mahalo-subaru_2013}. 
With deep NB imaging observations targeting H$\alpha$ emission at $z=2.53$, 
$\sim$ 100 H$\alpha$ emitters (HAEs) are identified as member galaxies 
associated with this protocluster so far 
\citep{hayashi_star-bursting_2012,hayashi_enhanced_2016,shimakawa_mahalo_2018}. 

Member galaxies were selected based on $r^{\prime}JK$ and $Br^{\prime}K_{s}$ color-color selection 
\citep{shimakawa_mahalo_2018}.
So far, 43 HAEs were spectroscopically confirmed 
with the rest-frame optical emission lines and/or CO(3--2) line 
\citep{shimakawa_identification_2014,shimakawa_correlation_2015,tadaki_environmental_2019}.

\begin{figure}%[H]
 \centering
 \includegraphics[keepaspectratio, scale=0.85]
      {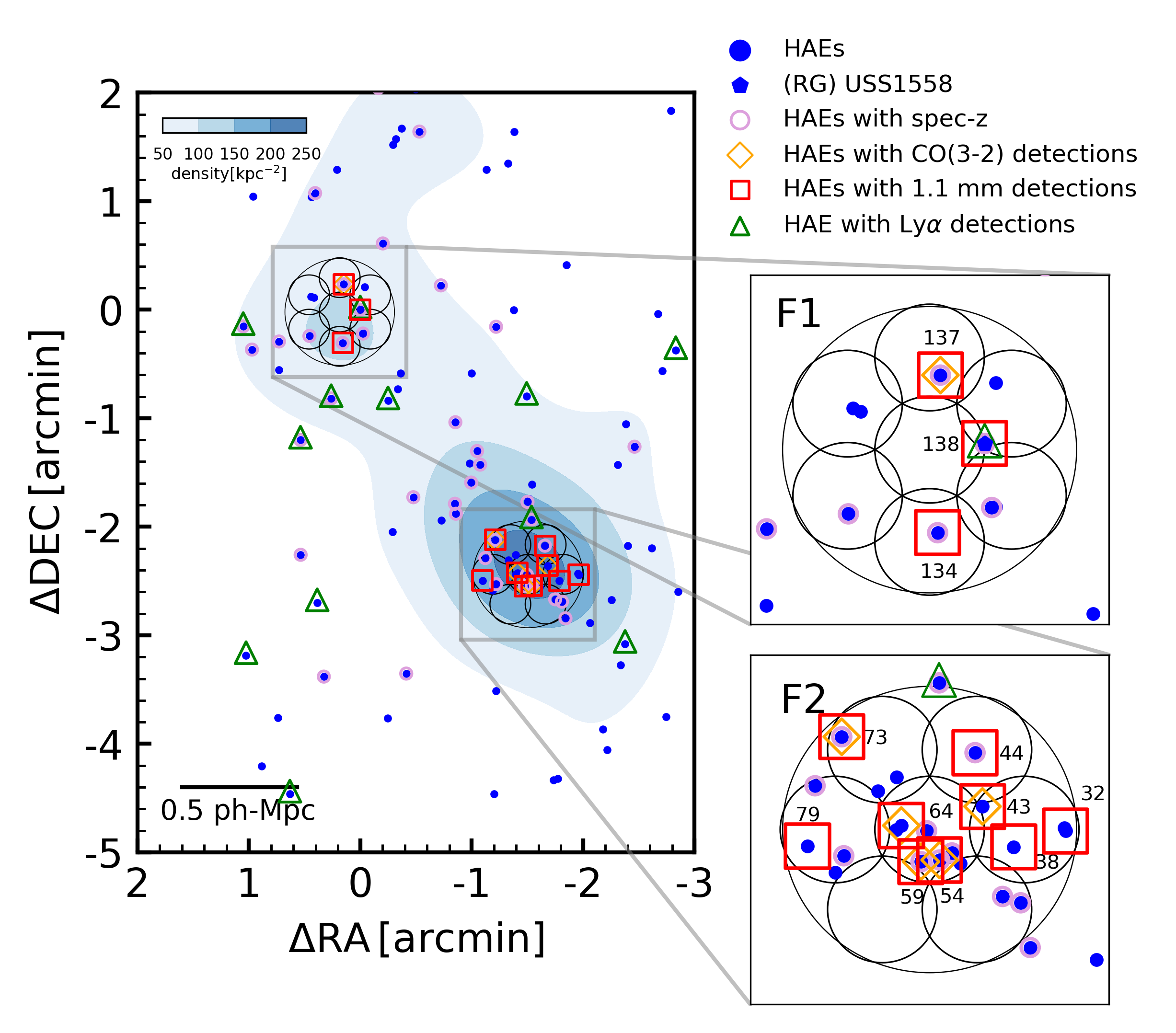}
 \caption{
 The two-dimensional sky map of HAEs in USS1558 and the close-up views of the two dense groups, 
 namely F1 and F2, targeted by our ALMA observations. 
 The coordinates are shown with respect to the location of RG.
 HAEs detected with 1.1~mm are tagged by their source IDs.
 The projected number density of HAEs is estimated by 2-D kernel density estimation 
 with $\sigma=0.5$ arcmin and shown as a contour map.
 %Purple open circles show HAEs with spectroscopic redshifts 
 %from the near-infrared spectroscopy \citep{shimakawa_identification_2014,shimakawa_correlation_2015}.  
 %open green triangles show HAEs with Ly$\alpha$ detections \citep{shimakawa_direct_2017}, 
 %open orange squares show HAEs with CO(3-2) detections \citep{tadaki_environmental_2019}, 
 %and open red squares show HAEs with Band-6 detections from this study.
 In the panels showing the close-up views of the two dense groups, 
 small and big circles represent the field coverage of ALMA Band-6 and Band-3 observations, respectively.
 }
 \label{fig:fov}
\end{figure}

Figure \ref{fig:fov} shows the spatial distribution of HAEs in USS1558. 
HAEs appear to distribute along a filamentary structure and to form several dense groups 
\citep{hayashi_star-bursting_2012}. 
The densest group of HAEs is located at $\sim1.5$~ph-Mpc away from RG to the southwest.
The number density of HAEs in the densest group is $\sim20$ times higher 
than that of the general fields. 
USS1558 is one of the densest star-forming protoclusters discovered to date at $z\sim2$. 
We also see a spatial concentration of HAEs in the vicinity of RG (Figure~\ref{fig:fov}). 
Hereafter, we call 
the dense group of HAEs in the immediate vicinity of RG 
and the densest group of HAEs Field-1 (F1) and Field-2 (F2), respectively 
(Figure~\ref{fig:fov}). 
The mean separation of HAEs in F1 and F2 is $<200$ and  $<100$ ph-kpc, respectively 
\citep{shimakawa_mahalo_2018}. 
Such a close separation is enough to overlap individual halos of HAEs, 
which means that some kind of physical interaction is expected.  
Given the group-like structure and no diffuse X-ray emission detection in this field, 
USS1558 is considered to be an immature system at the stage of vigorous assembly. 
One of the notable features of USS1558 is that 
the HAEs in the two dense groups (F1 and F2) have systematically 
high SFRs than those in the intergroup regions \citep{shimakawa_mahalo_2018}. 

In order to further investigate the origin of 
the elevated star-forming activities in USS1558 
and the environmental effects on the gas properties of the member galaxies, 
we conducted ALMA Band-6 observations for the two dense cores in the protocluster. 
In this field, we have conducted ALMA Band-3 observations 
targeting CO(3--2) emission lines in Cycle 3
(2015.1.00395.S; PI: T. Kodama; \citealt{tadaki_environmental_2019}).

\subsection{ALMA Band-6 observations}
\label{subsec:almadetection}

\begin{figure}[tb]
 \centering
 \includegraphics[keepaspectratio, scale=0.7]
      {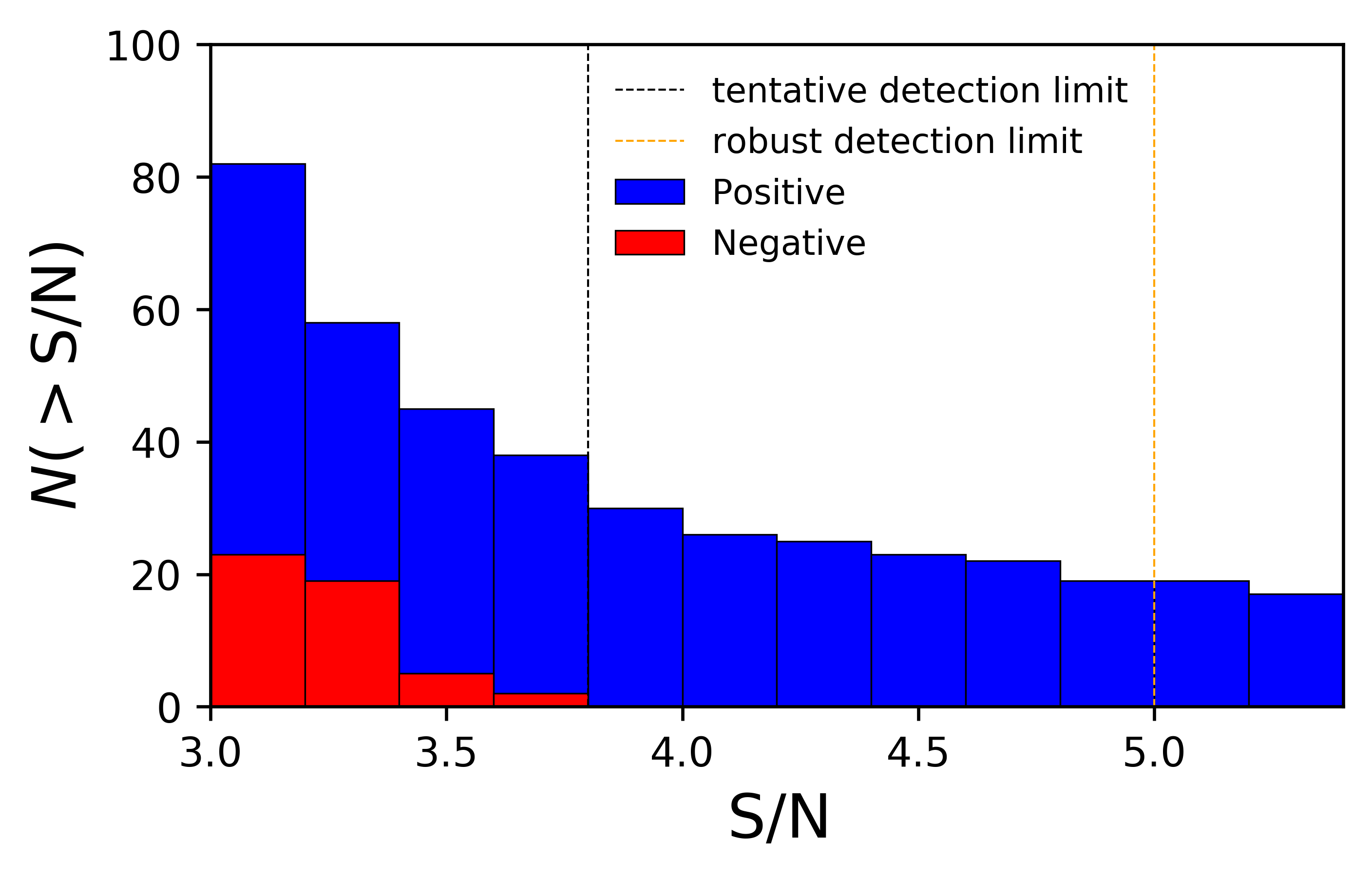}
 \caption{
 Cumulative number counts of positive (blue histogram) and negative (red) peaks 
 detected in the ALMA 1.1~mm maps as a function of S/N. 
 We set $3.8\sigma$ or $5.0\sigma$ 
 as a tentative or robust detection limit, respectively.
 }
 \label{fig:number_count}
\end{figure}

\begin{figure*}[!tb]
 \centering
 \includegraphics[width=\textwidth]
      {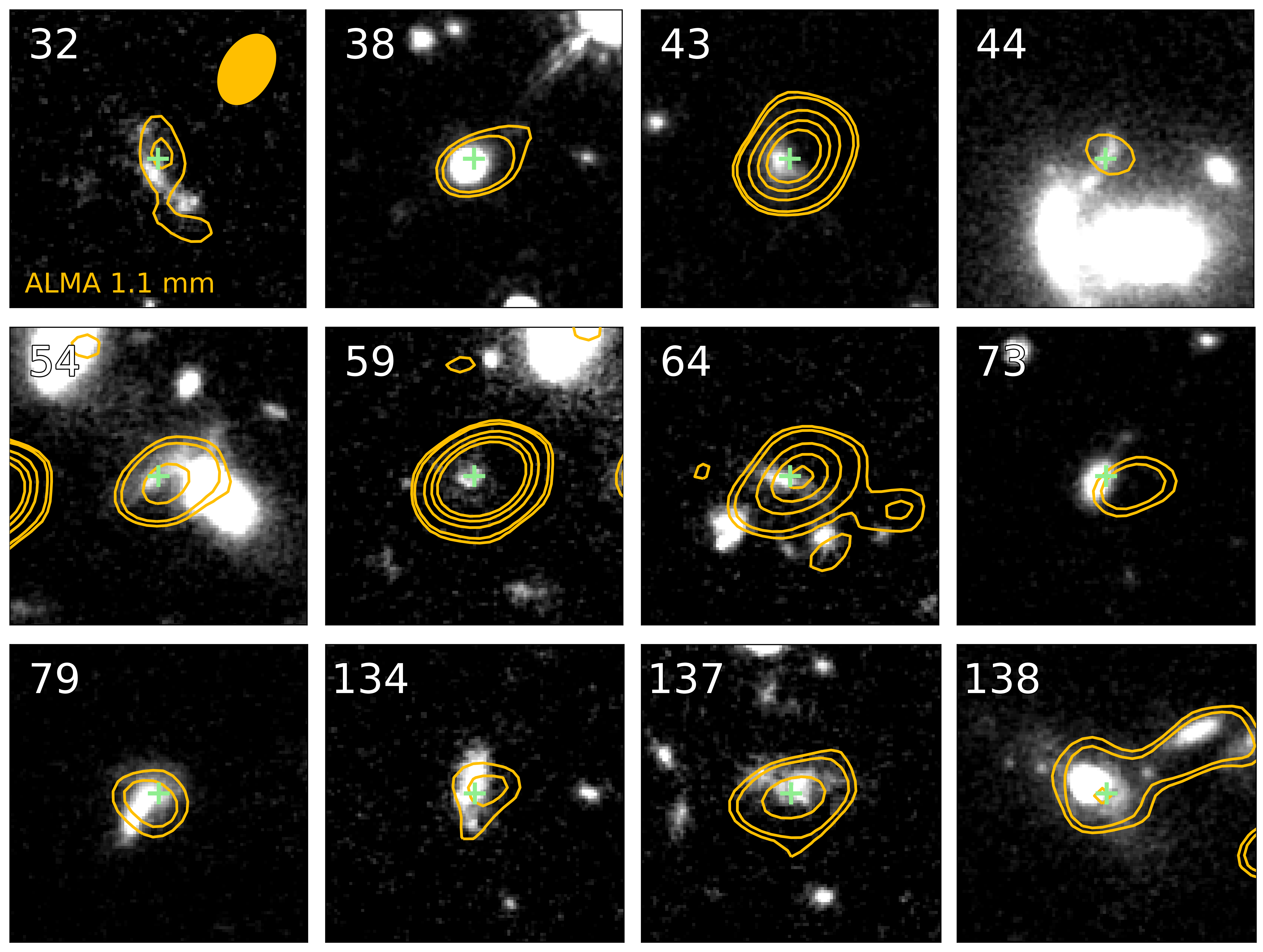}
 \caption{
 The $6'' \times 6''$cutout {\it HST}/WFC3 F160W images of the individual HAEs 
 with 1.1 mm detections. 
 Orange contours present $[3,4,8,12,16]\times\sigma$ in the ALMA 1.1~mm maps.
 %Their IDs are shown at the top left corner of the images.
 A cross mark shows the position of each HAE in the NB image.
 The beam size is shown in the top left panel.
 }
 \label{fig:HAE_ALMAdetected}
\end{figure*}

ALMA Band-6 observations were performed in Cycle5 (2017.1.00506.S, PI: T. Kodama) 
with a total on-source time of $\sim 8.7$ hours 
for 14-pointing mosaic observation (Figure~\ref{fig:fov}). 
We observed the two dense groups in USS1558, namely, F1 and F2. 
Four spectral windows with an effective width of 1.875 GHz are set to observe continuum emission at the central frequency of 265 GHz ($\sim 1.1 \rm mm$). 
The data was taken in the Time Division Mode (TDM).

The data calibration was done using the Common Astronomy Software Application Package 
\citep[{\sc casa};][]{mcmullin_casa_2007} version 5.4.0.
The continuum maps were obtained by a {\sc casa} task {\sc tclean} 
with natural weighting of the visibilities to obtain the highest sensitivity.
The beam size of the constructed maps is ${1''.51} \times 1''.07$ at a position angle of $-66$ deg. 
The root-mean-square (RMS) error noise is $0.020\, \rm mJy\,beam^{-1}$.
Since a typical size of the member galaxies is smaller than the beam size \citep{shimakawa_mahalo_2018}, 
we consider that the integrated flux density is the same as the peak flux 
after primary beam correction.

In order to determine a detection threshold, 
we count the number of positive/negative peaks as a function of the signal-to-noise ratio (S/N), as shown in Figure \ref{fig:number_count}.
No negative peaks below $3.8\sigma$ were found. 
It is suggested that, however, there is a systematic overestimation of fluxes due to the effect of noise and confusion called ``flux boosting'' \citep{scott_scuba_2002, umehata_alma_2017}.
Since the flux boosting effects become non-negligible at lower S/N, 
we regard $>5\sigma$ detections as robust detections and 3.8--5$\sigma$ detections as tentative detections in the following analyses. 
In total, 19 and 11 sources were detected at $>5\sigma$ and 3.8--5$\sigma$, respectively. 
%We set $3.8\sigma$ as a tentative detection threshold 
%because no peaks below $-3.8\sigma$ were found.
%It is suggested that there is a systematic overestimation of fluxes 
%due to the effect of noise and confusion called ``flux boosting'' 
%\citep{scott_scuba_2002, umehata_alma_2017}.
%Since the flux boosting effects become non-negligible at lower S/N, 
%we can not ignore the effects of flux boosting at lower S/N, 
%we set $5.0\sigma$ as a robust detection threshold. 
%In total, 30 and 19 sources were detected at $>3.8\sigma$ and $>5.0\sigma$, respectively.
%Then, we cross-matched these ALMA sources with the HAEs in the two fields 
%with a searching radius of $0''.6$. 
As a result, we obtained nine HAEs with robust ALMA detection 
and three with tentative ALMA detection. 
The images of the 1.1~mm-detected HAEs are shown in Figure~\ref{fig:HAE_ALMAdetected}.

As for 18 1.1~mm sources left,
ten sources have counterparts 
in the UV-to-optical images.
None of the ten sources with the UV/optical counterparts 
%are detected in NB images and 
satisfied the selection criteria of NB emitters or 
the color-color selections  
($r'JK_{s}$ and $Br'K_{s}$) 
used to select H$\alpha$ emitters at $z=2.53$ in \citet{shimakawa_mahalo_2018}.
Therefore these ten sources are likely to be  
foreground/background sources.

\begin{figure}
    \centering
    \includegraphics[width=0.6\columnwidth]{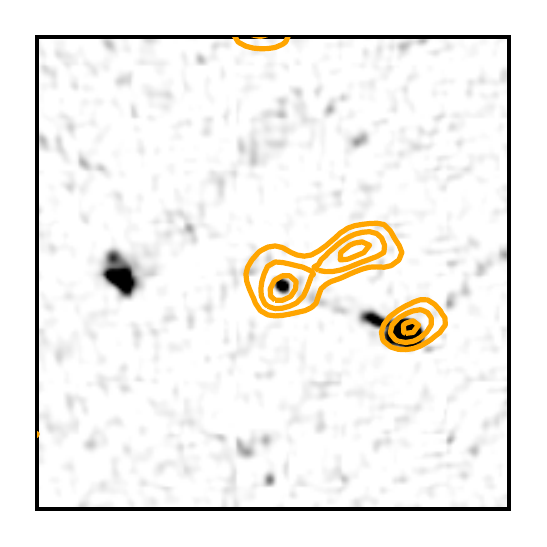}
    \caption{VLA 8.44~GHz image centered at RG (image size: 13~arcsec$\times$13~arcsec). 
    The contours show [3, 5, 7]$\times \sigma$ in the ALMA 1.1~mm map.  
    We can see that a bright 1.1~mm continuum emission is spatially associated to a radio emission, which is called the southern hot spot in \citet{pentericci_vla_2000}. 
    }
    \label{fig:vlamap_rg}
\end{figure}

Out of eight sources that do not have UV/optical counterparts,
one source is located at $\sim4$~arcsec away from the RG (a separate source located toward the right edge). 
Figure~\ref{fig:vlamap_rg} shows the VLA 8.44~GHz image around the RG taken from the NRAO Science Data Archive\footnote{\url{https://archive.nrao.edu/archive/archiveimage.html}}. 
We found that this 1.1~mm continuum source is spatially associated to a radio emission detected with VLA, which is called the southern hot spot in \citet{pentericci_vla_2000}.
Therefore it likely corresponds to a jet ejected from the RG, and its 1.1-mm flux is probably dominated by the synchrotron emission from the radio jet. 

%one source near RG turned out to be dust emission from the jet from RG.
%Its coordinate in the 1.1~mm map is matched with 
%that of the radio jet detected in VLA observations by 
%\citet{pentericci_vla_2000}.
%\color{black}

Three sources among the 1.1~mm continuum sources without UV/optical counterpart are detected in ALMA 3~mm continuum maps (2016.1.00461.S; PI: M.\ Hayashi). 
%which means that they are real sources. 
We compared the photometries from UV-to-sub-mm
(upper limit values are used for the UV-optical bands)
with the average SED of ALESS sources
with $A_{\rm V} > 3$~mag \citep{da_cunha_alma_2015}
by changing redshifts as done in \citet{yamaguchi_alma_2019}. 
The three sources are expected to be dusty sources 
at higher redshifts ($z>3$) from this comparison. 
As for the remaining four sources with no UV/optical counterpart, 
S/N of their 1.1~mm continuum fluxes is less than 5. 
Given their relatively low S/N, these four sources with no UV/optical counterpart may be false detections.

To summarize, in the 1.1~mm dust continuum maps, 
we find no additional plausible member galaxies except for HAEs. 
In the following analyses, we focus on nine HAEs with a robust detection and three HAEs with a tentative detection at 1.1~mm. 
As mentioned above, the 1.1~mm continuum fluxes of the HAEs with a tentative detection might be slightly overestimated due to the flux boosting effect. 
%We focus on 12 HAEs with
%$> 3.8\sigma$ detection at 1.1~mm 
%in the following analyses. 
%color{red}
%We caution the reader that our sample includes 3 HAEs with tentative ALMA detections to enlarge the sample size.
%\color{black}
The measured 1.1 continuum fluxes of the HAEs after primary beam correction are summarized in Table~\ref{tab:summary}.

%We also conducted aperture photometry at the positions of the HAEs in the 1.1~mm maps
%to check the validity of our flux measurement based on the peak flux.
%We set a $1''.2$ radius ($\sim2\times$ beam size) and 
%measure the flux using a {\sc casa} task {\sc imstat}.
%RMS is calculated by running {\sc imstat} at random positions in the 1.1~mm maps. 
%The aperture fluxes are consistent with the peak fluxes within the margin of error, which means that 
%the flux measurement based on the peak flux seems to be valid.

We conducted some tests to check the validity of the flux measurements based on the peak flux.
In the first place, we measured aperture photometry at the positions of the HAEs in the 1.1~mm continuum maps. 
We set a $1''.2$ radius ($\sim2\times$ beam size) and 
measure the flux using a {\sc casa} task {\sc imstat}.
RMS is calculated by running {\sc imstat} at random positions in the 1.1~mm maps. 
The aperture fluxes are consistent with the peak fluxes within the margin of error, which means that the flux measurement based on the peak flux seems to be valid. 
Some of the HAEs (HAE 32, 64, and 138), however, have nearby companions with S/N $>3$ in the 1.1~mm continuum maps as shown in Figure~\ref{fig:HAE_ALMAdetected}. 
The peak flux of the three HAEs might be contaminated by the continuum emission from their companions. 
We measured their fluxes by fitting a 2D Gaussian to the HAEs and their companions simultaneously with a {\sc casa} task {\sc imfit}. 
As for the HAEs, the positions were fixed to the coordinates measured in the NB image. 
As for the companions, the positions were fixed at the flux peak in the 1.1~mm  maps. 
The peak fluxes measured with {\sc imfit} of the three HAEs turned out to be $\sim$10--30~\% smaller than the peak fluxes measured as a single galaxy. 
However, we confirmed that the gas to stellar mass ratios and star-formation efficiencies estimated from the two fluxes agree with each other within the uncertainties, which means that our final results are not affected by the flux contamination from nearby companions.

\subsection{Physical quantities from ancillary data}

Multi-wavelength photometries from UV to near-infrared (NIR) 
are available in USS1558. 
In addition to broad-band/NB images ($B, r^{\prime}, z^{\prime}, J, H, K_{s}$, 
and NB2315 with Suprime-Cam and  MOIRCS on Subaru), 
$F160W$ images taken with {\it the Hubble Space Telescope (HST)}/WFC3 are available  \citep{hayashi_enhanced_2016}. 
With the multi-wavelength data, the fundamental properties of galaxies, such as stellar mass, star formation rate (SFR), and effective radius ($R_{\rm e}$),  
have been already derived in the previous studies 
\citep{hayashi_star-bursting_2012,hayashi_enhanced_2016,shimakawa_mahalo_2018}.
We briefly introduce how to derive them in the following sections.

\subsubsection{Stellar mass and SFR measurement}
\label{subsubsec:SFR measurement}

The stellar masses and SFRs of HAEs were estimated by \citet{shimakawa_mahalo_2018}. 
The stellar mass was derived with the Spectral Energy Distribution (SED) fitting code, 
{\sc fast} \citep{kriek_ultra-deep_2009}. 
They used $B, r^{\prime}, z^{\prime}, J, H, F160W, K_{s}$-band photometries for the SED fitting.
The SFR was derived from the H$\alpha$ NB fluxes 
with the \citet{kennicutt_star_1998} prescription 
assuming the \citet{chabrier_galactic_2003} IMF.
The dust extinction is corrected for with the \citet{calzetti_dust_2000}  
attenuation law and $A_{V}$ from {\sc fast} 
assuming $E(B-V)_{\rm nebular} = E(B-V)_{\rm stellar}$ 
\citep[e.g.,][]{erb_h-alpha_2006,reddy_dust_2010,reddy_mosdef_2015}.

\begin{figure}[t]
 \centering
 \includegraphics[keepaspectratio, scale=0.8]
      {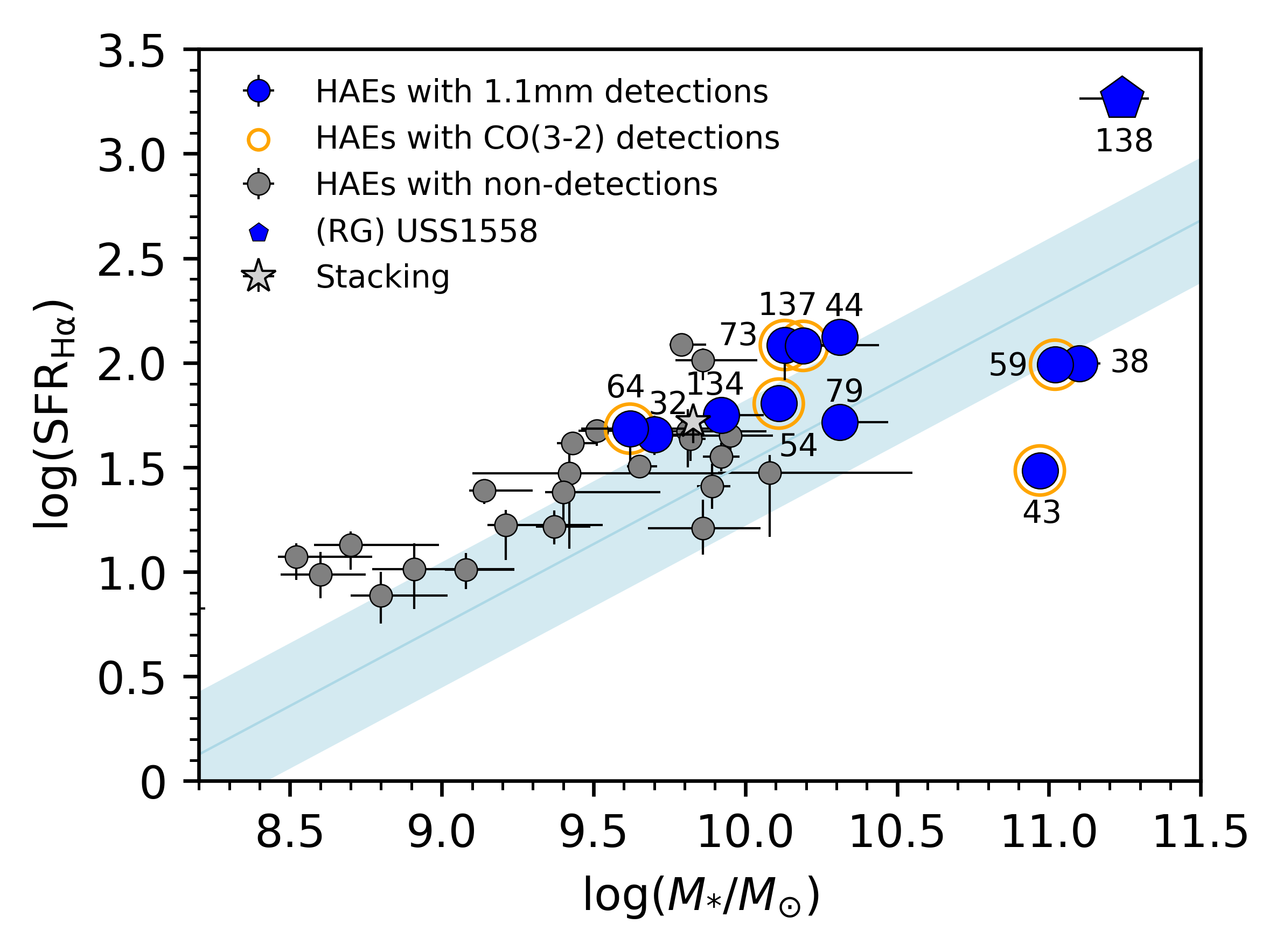}
 \caption{
 $M_{*}$--SFR relation for HAEs in F1 and F2. 
 The solid line represents the star-forming MS at $z=2.5$ 
 from \citet{speagle_highly_2014}. 
 A shaded region shows the range of $\pm 0.3$ dex from the MS, 
 corresponding to a typical $1\sigma$ scatter of the MS.
 The HAEs in the dense cores are located $\sim0.3$~dex above the MS at $z=2.5$ systematically. 
 %systematically higher SFRs than those in a less dense region
 %\citep{shimakawa_mahalo_2018}.
 }
 \label{fig:M_SFR}
\end{figure}

Figure \ref{fig:M_SFR} shows the relation between stellar mass and SFR 
for the HAEs in the two dense groups of USS1558.
As shown in \citet{shimakawa_mahalo_2018}, 
the majority of the HAEs in the dense groups are located $\sim0.3$~dex above 
the $M_{*}$--SFR relation 
(the so-called main sequence (MS) of star-forming galaxies)
at $z=2.5$ from \citet{speagle_highly_2014}. 
We note that \citet{chabrier_galactic_2003} IMF is consistently
assumed here 
(The original $M_{*}$--SFR relation in \citet{speagle_highly_2014} assumed the \citet{kroupa_imf_2001} IMF).
RG (ID138) is an only X-ray source with $L_{X}>2\times 10^{43}\ \rm erg\,s^{-1}$ 
in these regions \citep{macuga_fraction_2019}.
Note that the SFR of RG from H$\alpha$ luminosity may be overestimated 
due to the contribution from the active galactic nuclei (AGN). 

%In order to check the robustness of our H$\alpha$-based SFR, 
We also estimated SFRs using a different SED fitting code, 
{\sc magphys} \citep{da_cunha_simple_2008,da_cunha_alma_2015}. 
We used ancillary photometry as mentioned above and 1.1~mm continuum flux for SED fitting. 
We confirmed that the SFR from {\sc magphys} are consistent with 
the H$\alpha$-based SFR at least for the HAEs located above the main sequence, 
which means that the enhanced star-formation activities observed 
in the dense cores seem to be valid.

\subsubsection{Size measurement}
The half-light radii of the HAEs were estimated by \citet{shimakawa_mahalo_2018}.
They measured the size from {\it F160W} images by using 
{\sc galfit} the version 3.0 code \citep{peng_detailed_2010}.
They use {\sc SExtractor} \citep{bertin_sextractor_1996} to get 
the initial parameters needed for fitting.
As for HAE44 and HAE54, half-light radii were not estimated 
in the previous study 
because these two HAEs are close to bright sources. 
We measured the half-light radii of the two HAEs 
by masking the nearby bright sources with a segmentation map 
from {\sc SExtractor}. 

We note that the estimated size of HAE59 
($R_{\rm e}=14.55\pm4.83$~kpc)
is a factor of $\sim$3 larger
as compared to the mass-size relation of 
star-forming galaxies at similar redshifts \citep{van_der_wel_3d-hstcandels_2014}. 
Given that this source is relatively faint in the {\it F160W} image with $m_{F160W}=24.2$~mag as compared to the other HAEs, 
%This source is relatively faint in the {\it F160W} image. 
{\sc galfit} might not be able to well fit the structural properties of this source, and thus, overestimate its size.

\section{Analysis}
\label{sec:ana}

\subsection{Gas mass estimation}
\label{subsec:gas mass estimation}

As for HAEs detected 
with both dust continuum and CO(3--2) line, 
we derive molecular gas masses with the two tracers 
to check the consistency between them (Section~\ref{subsec:coexcitation}). 
We describe how to derive the molecular gas mass 
with each tracer below.

\subsubsection{CO-based gas mass}
\label{subsubsec:cogasmass}
We derived molecular gas masses from the CO(3--2) emission line 
as done in \citet{tadaki_environmental_2019}.
We used CO(3--2) line fluxes measured by \citet{tadaki_environmental_2019} 
but assumed different line ratios between CO(3--2) and CO(1--0) as discussed in Section~\ref{subsec:coexcitation}. 
Molecular gas mass considering a metallicity-dependent conversion factor of $\alpha_{{\rm CO}J}$ is derived as follows:

\begin{eqnarray}
&M_{\rm mol, CO}/M_{\odot} = \alpha_{{\rm CO} J}\times L'_{{\rm CO}J}\\
&\alpha_{{\rm CO} J} = \alpha_{\rm CO 1} \times \frac{1}{r_{J1}} \times f_{\rm corr, CO} \\
&\,\,\,\,\,\,\,\,\,\,\,\,\,\,\,\,\,\,\,\,\,\,\,\,\,\,\,\,(M_{\odot}\ ({\rm K\,km\,s^{-1}\,pc^{2}})^{-1})\nonumber\\
&f_{\rm corr, CO} = \sqrt{\begin{aligned}0.67\exp({0.36\times10^{-(\log(\rm 12+O/H)-8.67)}}) \\
\times 10^{-1.27(\log(\rm 12+(O/H)-8.67))}\end{aligned}},\,\,\, 
\end{eqnarray}

where $\alpha_{{\rm CO}J}$ is a conversion factor of 
$J$-to-($J$--1) transition CO line luminosity to a molecular gas mass, 
$r_{J1} = L^{\prime}_{{\rm CO}(J-(J-1))}/L^{\prime}_{{\rm CO}(1-0)}$, 
$\alpha_{\rm CO1}=4.3$, and 
$f_{\rm corr, CO}$ is the correction factor of 
metallicity dependence of $\alpha_{{\rm CO}J}$ \citep{tacconi_phibss_2018}.
Here, $12+\log(\rm O/H)$ is the gas-phase metallicity 
measured with [NII]/H$\alpha$ ratio 
and a calibration relation by \citet{pettini_o_2004}.
The gas-phase metallicities for the CO(3--2)-detected HAEs 
are inferred from their stellar masses and 
the mass--metallicity relation in \citet{genzel_combined_2015-1} (Equations 12a). 
The metallicity correction factor for galaxies 
with $\log(M_{*}/M_\odot)=9.5, 10$ and  $10.5$ 
is $f_{\rm corr, CO}=4.3,2.3$, and $1.5$, respectively. 
In \citet{tadaki_environmental_2019}, 
the line ratio between CO(3--2) and CO(1--0), $r_{\rm 31}$, 
is assumed to be 0.56 \citep{genzel_metallicity_2012}. 
$r_{\rm 31}$ values for star-forming galaxies at high redshifts 
reported in the literature show a large ($\sim$ 0.2 dex) variety 
\citep[e.g.,][]{aravena_co1-0_2014,daddi_co_2015,sharon_total_2016,riechers_vla-alma_2020}.
On top of that, a variation in $\alpha_{\rm CO1}$ yields $\sim0.15$ dex uncertainties \citep{genzel_metallicity_2012}.
We will discuss the assumption of $r_{\rm 31}$ for the HAEs in USS1558 in Section~\ref{subsec:coexcitation}. 
Considering these errors, and add them in quadrature, the measurement of CO-based molecular gas mass yields 0.25~dex of systematic error in total.  
%We will discuss $r_{\rm 31}$ values for the HAEs in USS1558  
%in Section~\ref{subsec:coexcitation} by comparing 
%the CO(3--2)-based molecular gas masses 
%with those estimated with the dust continuum emission.
%, and 
%the scatter of mass-metallicity relation yields $\sim0.1$ dex uncertainties \citep{elbaz_reversal_2007-1}. 

\subsubsection{Dust-based gas mass}
\defcitealias{scoville_ism_2016}{S16}

We derived molecular gas masses with the Rayleigh-Jeans (RJ) dust continuum method 
presented in \citet[][hereafter \citetalias{scoville_ism_2016}]{scoville_ism_2016}.
The dust emission at RJ-tail ($\lambda_{\rm rest} \gtrsim 250 \rm um$) 
can be used 
as a probe of global cold gas mass.
\citetalias{scoville_ism_2016} empirically calibrated the conversion factor 
from dust continuum luminosity at RJ-tail to gas mass using 145 galaxies, 
including star-forming galaxies and starburst galaxies at low redshifts 
and sub-millimeter bright galaxies (SMGs) at $z=$2--3. 
The molecular gas mass is given by 

\begin{eqnarray}
\label{eq:sco}
M_{\rm mol,\,S16}=1.78\, S_{\nu_{\rm obs}}{\rm [mJy]}\,(1+z)^{-(3+\beta)}\nonumber\\
\times \left( \frac{\nu_{\rm 850\mu m}}{\nu_{\rm obs}}\right)^{2+\beta}\times(d_{L}[{\rm Gpc}])^2\nonumber\\
\times \left\{ \frac{6.7\times10^{19}}{\alpha_{\rm 850}} \right\} \, \frac{\Gamma_{0}}{\Gamma_{\rm RJ}}\, 10^{10}M_{\odot}\nonumber\\
{\rm for} \, \, \lambda_{\rm rest} \gtrsim 250\, \mu {\rm m},
\end{eqnarray}
where $d_L$ is the luminosity distance 
and $\alpha_{850} = 6.7\times 10^{19}\,{\rm erg\, s^{-1}\, Hz^{-1}\,}M_{\odot}^{-1}$. 
The dust spectral index $\beta$ is assumed to be 1.8. 
The correction term from RJ to Plank function $\Gamma_{\rm RJ}$ is given by 
\begin{equation}
\Gamma_{\rm RJ}(T_{\rm d}, \nu_{\rm obs}, z) = \frac{h\nu_{\mathrm obs}(1+z)/kT_{\rm d}}{\exp(h\nu_{\rm obs}(1+z)/kT_{\rm d}) - 1}, 
\end{equation}

\noindent
and $\Gamma_{0} = \Gamma_{\rm RJ}(T_{\rm d}, \nu_{850}, 0)$. 
The dust temperature $T_d$ is assumed to be $25\ {\rm K}$, 
a typical value of the mass-weighted temperature in the local 
star-forming galaxies \citepalias{scoville_ism_2016}.

\citetalias{scoville_ism_2016} selected massive galaxies 
with $M_{*} > 2\times10^{10}\ M_\odot$, 
which have similar gas-phase metallicities as the solar value, 
for their calibration. 
Their sample galaxies are expected to have 
an almost constant dust-to-gas mass ratio of 
$\delta_{dg}=M_{\rm dust}/M_{\rm mol} \sim 1/100$. 
The gas-to-dust mass ratio is included in the conversion factor. 
Because our sample consists of less massive galaxies 
than those used in the calibration by \citetalias{scoville_ism_2016}, 
we tried to consider the metallicity effects of this method further. 
As in Section~\ref{subsubsec:cogasmass}, 
we infer the gas-phase metallicity from the mass--metallicity relation presented by \citet{genzel_combined_2015-1}. 
We then use the relation between the dust-to-gas mass ratio ($\delta_{dg}$) 
and the gas-phase metallicity given in \citet{genzel_combined_2015-1} 
in order to correct for the metallicity dependence as follows: 

\begin{eqnarray}
\delta_{dg} &=& 10^{-2 + 0.85\times((12+ \log(\rm O/H))-8.67)}, \\
M_{\rm mol,\,dust} &=& f_{\rm corr, dust} \times M_{\rm mol,\,S16} \\
&=& 10^{-0.85\times((12+ \log(\rm O/H)-8.67))} \times M_{\rm mol,\,S16}.\,\,\,\,\,\,\,\,
\end{eqnarray}

The metallicity correction factor for galaxies 
with $\log(M_{*}/M_\odot)=9.5, 10$ and  $10.5$ is 
$f_{\rm corr, dust}=3.3,2.1$, and $1.5$, respectively. 

We assume that the HAEs in USS1558 follow 
the mass--metallicity relation for field galaxies at similar redshifts. 
How the mass--metallicity relation of star-forming galaxies 
depends on the environments at $z\sim2$ has been discussed,   
and it is not conclusive yet \citep[e.g.,][]{valentino_metal_2015,shimakawa_early_2015,kacprzak_absence_2015, chartab_mosdef_2021}. 
\citet{shimakawa_early_2015} investigated 
the average gas-phase metallicities 
of HAEs in USS1558 by conducting stacking analyses, 
and they only gave the upper limits to the metallicities 
for the HAEs with $\log(M_{*}/M_{\odot})\lesssim10.3$. 
It is still unclear whether the HAEs in USS1558 
have a different mass--metallicity relation 
or follow the same relation as field galaxies. 
Therefore, we have decided to use the mass--metallicity relation 
for field galaxies.

When we use the mass--metallicity relation for protocluster galaxies 
shown in \citet{shimakawa_early_2015},
molecular gas masses become 0.2 and 0.1~dex lower 
for galaxies with $\log(M_{*}/M_\odot)=10.0$ 
and $10.5$, respectively. 
We note that our results are not significantly affected by the assumed mass-metallicity relations when taking into account the uncertainties on our gas mass measurements and the scatter of the field gas scaling relations (Section~\ref{subsec:scaling}). 
We would need individual metallicity measurements for more precise metallicity correction. 
We note that we do not consider 
the flux boosting effect on our 
1.1~mm continuum flux measurement 
(Section~\ref{subsec:almadetection}). 
The estimated molecular gas masses of the two sources with S/N$=$ 3.8--5.0 (HAE~32, 44, and 134)
maybe overestimated due to the flux boosting effect.

The uncertainty on the conversion factor $\alpha_{850\mu \rm m}$ 
in \citetalias{scoville_ism_2016} 
causes $\sim0.1$~dex uncertainty on the estimated 
molecular gas masses. 
Since the mass--matallicity relation has a scatter of $\pm 0.1$ dex as well, 
our molecular gas measurement based on the dust continuum emission 
has a systematic uncertainty of $\pm0.13$ dex in total.
We also note that, 
because the impact of AGN contribution peak in mid-IR but trails off in longer wavelength, 
the molecular gas estimation is not suffered significantly from the AGN contribution 
\citep{kirkpatrick_goods-herschel_2012, brown_cold_2017}.

\subsection{Stacking analysis}

\begin{figure}[t]
 \centering
 \includegraphics[width=0.48\textwidth]
      {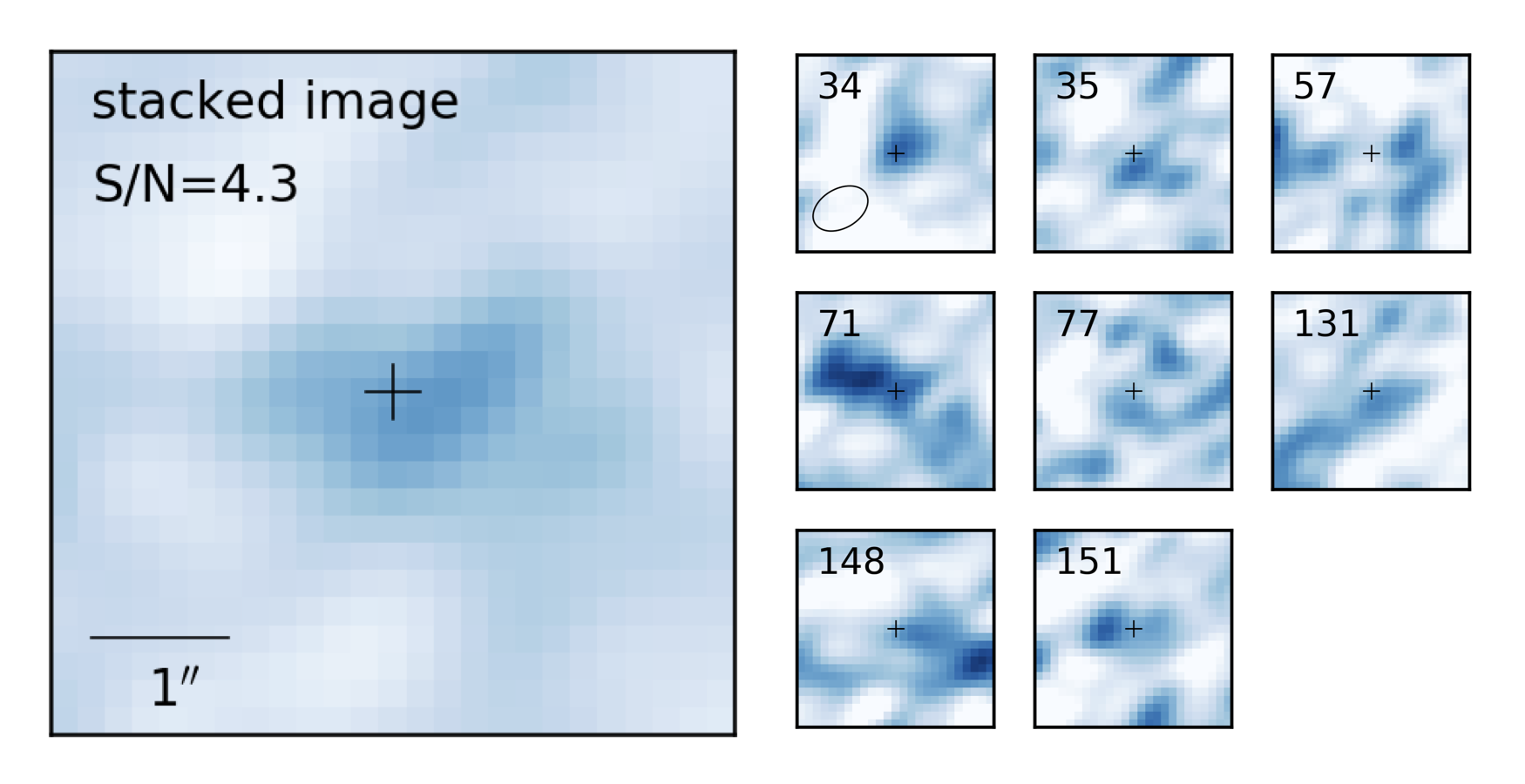}
 \caption{
 (Left) The stacked image of HAEs that are not detected individually 
 in Band-6 with $9.5 < \log(M_{*}/M_{\odot}) < 10.0$. 
 (Right) The $5'' \times 5''$cutout images of individual HAEs 
 used to make the stacked image. 
 Their IDs are shown at the top left corner of the images.
 Cross marks show the positions of individual HAEs 
 determined in the Subaru/MOIRCS NB image.
 The beam size is shown in the top left panel.
 }
 \label{fig:stack}
\end{figure}

As shown in Figure~\ref{fig:M_SFR}, 
the detection rate is high (9/10) for massive galaxies 
with $\log(M_{*}/M_{\odot})>10.0$ 
but low for less-massive galaxies (3/12) 
with $9.5 < \log(M_{*}/M_{\odot}) <10.0$.
In order to 
investigate the average gas properties of less massive galaxies, 
we conducted a stacking analysis 
for less massive galaxies.
We choose the non-detected HAEs with $9.5<\log(M_{*}/M_{\odot})<10.0$.
We excluded one HAEs contaminated by nearby bright sources 
in Band-6 continuum maps.
We stacked the cutout images of the remaining eight HAEs 
at the position determined in the Subaru/MOIRCS NB image.
Figure \ref{fig:stack} shows the stacked image and the cutout images of the HAEs 
used for stacking. 
The stacked image has $4.3\sigma$ significance 
at the center, which satisfies our detection criterion 
(Section~\ref{subsec:almadetection}).

%%%%%%%%%%%%%%%%%%%%%%%%%%%%%%%%%%%%%%%%%%%%%%%%%%%%%%%%%%%%%%%%%%%%%%%%%
\movetableright=0.002in
\begin{rotatetable*}
\begin{deluxetable*}{lccccccccccc}
\tablecaption{Physical quantities of HAEs in USS1558 detected in Band-6 observations}
\tablewidth{400pt}
\tabletypesize{\scriptsize}
\tablehead{
\colhead{$\rm ID$\tablenotemark{a}} & 
\colhead{$\rm RA$\tablenotemark{a}} & 
\colhead{$\rm DEC$\tablenotemark{a}} & 
\colhead{$\log(M_{*}/M_{\odot})$\tablenotemark{a}} & 
\colhead{$\rm \log(SFR_{H_{\alpha}})$\tablenotemark{a}} & 
\colhead{$\rm \log(\rm SFR_{SED})$\tablenotemark{a}} & 
\colhead{$R_{\rm e}$\tablenotemark{a}} &
\colhead{$S_{\rm 1.1mm}$\tablenotemark{b}} &
\colhead{$\log(M_{\rm mol,\,dust}/M_{\odot})$} &
\colhead{$\log(M_{\rm mol,\,CO}/M_{\odot})$\tablenotemark{c}} &
\colhead{$\log(\mu_{\rm mol})$} &
\colhead{$\log(\rm SFE)$} \\
%%%%%%%%%%
\colhead{} & 
\colhead{(J2000)} &    
\colhead{(J2000)} &    
\colhead{} & 
\colhead{$[M_{\odot}/\rm yr]$} & 
\colhead{$[M_{\odot}/\rm yr]$} & 
\colhead{$[\rm kpc]$} &
\colhead{$[\rm mJy]$} &
\colhead{} &
\colhead{} &
\colhead{} &
\colhead{$[\rm Gyr^{-1}]$}}
%\decimalcolnumbers
\startdata
$32$\tablenotemark{d} & $240.289602$ & $-0.520247$ & $9.7^{+0.15}_{-0.13}$ & $1.66^{+0.09}_{-0.1}$ & $1.66^{+0.01}_{-0.41}$ & $4.59\pm0.08$ & $0.13$ & $10.6\pm0.07$ & $<11.14$ & $0.9^{+0.16}_{-0.15}$ & $0.06^{+0.11}_{-0.12}$ \\
$38$ & $240.292499$ & $-0.521166$ & $11.1^{+0.07}_{-0.06}$ & $2.0^{+0.04}_{-0.07}$ & $2.09^{+0.09}_{-0.1}$ & $1.47\pm0.24$ & $0.16$ & $10.3\pm0.05$ & $<10.55$ & $-0.8^{+0.09}_{-0.08}$ & $0.7^{+0.07}_{-0.09}$ \\
$43$ & $240.294239$ & $-0.518846$ & $10.97^{+0.05}_{-0.03}$ & $1.49^{+0.08}_{-0.07}$ & $1.87^{+0.0}_{-0.01}$ & $5.44\pm0.21$ & $0.53$ & $10.83\pm0.02$ & $10.91\pm0.07$ & $-0.14^{+0.05}_{-0.03}$ & $-0.35^{+0.08}_{-0.07}$ \\
$44$\tablenotemark{d} & $240.294654$ & $-0.515767$ & $10.31^{+0.05}_{-0.03}$ & $2.12^{+0.03}_{-0.03}$ & $1.93^{+0.13}_{-0.08}$ & $7.19\pm0.17$ & $0.08$ & $10.16\pm0.11$ & $<10.75$ & $-0.15^{+0.12}_{-0.12}$ & $0.96^{+0.11}_{-0.11}$ \\
$54$ & $240.296629$ & $-0.521909$ & $10.11^{+0.05}_{-0.02}$ & $1.8^{+0.04}_{-0.03}$ & $1.93^{+0.4}_{-0.0}$ & $3.87\pm2.11$ & $0.22$ & $10.68\pm0.04$ & $11.11\pm0.04$ & $0.57^{+0.06}_{-0.04}$ & $0.12^{+0.05}_{-0.05}$ \\
$59$ & $240.297669$ & $-0.522007$ & $11.02^{+0.0}_{-0.0}$ & $1.99^{+0.03}_{-0.03}$ & $2.21^{+0.28}_{-0.41}$ & $14.55\pm4.83$ & $1.0$ & $11.1\pm0.01$ & $11.18\pm0.02$ & $0.08^{+0.01}_{-0.01}$ & $-0.11^{+0.03}_{-0.03}$ \\
$64$ & $240.298762$ & $-0.519938$ & $9.62^{+0.27}_{-0.16}$ & $1.68^{+0.08}_{-0.16}$ & $1.7^{+0.01}_{-0.01}$ & $4.81\pm0.15$ & $0.37$ & $11.08\pm0.02$ & $11.31\pm0.08$ & $1.46^{+0.27}_{-0.16}$ & $-0.4^{+0.08}_{-0.16}$ \\
$73$ & $240.302086$ & $-0.514857$ & $10.13^{+0.31}_{-0.04}$ & $2.09^{+0.03}_{-0.17}$ & $2.16^{+0.0}_{-0.15}$ & $2.45\pm0.03$ & $0.17$ & $10.57\pm0.05$ & $11.02\pm0.07$ & $0.44^{+0.31}_{-0.06}$ & $0.51^{+0.06}_{-0.18}$ \\
$79$ & $240.303995$ & $-0.521118$ & $10.31^{+0.16}_{-0.04}$ & $1.72^{+0.05}_{-0.08}$ & $2.08^{+0.0}_{-0.0}$ & $4.75\pm0.03$ & $0.13$ & $10.39\pm0.07$ & $<10.88$ & $0.08^{+0.17}_{-0.08}$ & $0.33^{+0.08}_{-0.1}$ \\
$134$\tablenotemark{d} & $240.324903$ & $-0.484626$ & $9.92^{+0.14}_{-0.04}$ & $1.75^{+0.05}_{-0.07}$ & $1.46^{+0.01}_{-0.0}$ & $4.46\pm0.05$ & $0.09$ & $10.37\pm0.09$ & $<10.87$ & $0.45^{+0.17}_{-0.1}$ & $0.38^{+0.11}_{-0.12}$ \\
$137$ & $240.324746$ & $-0.475606$ & $10.19^{+0.16}_{-0.11}$ & $2.08^{+0.07}_{-0.09}$ & $1.95^{+0.01}_{-0.23}$ & $6.12\pm0.1$ & $0.24$ & $10.69\pm0.04$ & $10.92\pm0.05$ & $0.5^{+0.16}_{-0.12}$ & $0.39^{+0.08}_{-0.1}$ \\
$138$\tablenotemark{e} & $240.322277$ & $-0.479524$ & $11.24^{+0.09}_{-0.14}$ & $3.26^{+0.07}_{-0.07}$ & $1.4^{+0.0}_{-0.01}$ & $2.34\pm0.09$ & $0.19$ & $10.35\pm0.05$ & -- & $-0.89^{+0.1}_{-0.15}$ & $1.92^{+0.08}_{-0.08}$
\enddata
\tablecomments{
\tablenotetext{a}{From \citet{shimakawa_mahalo_2018}.}
\tablenotetext{b}{Corrected for the primary beam attenuation.}
\tablenotetext{c}{Derived from CO(3--2) line fluxes from \citet{tadaki_environmental_2019} using method described in Section~\ref{subsubsec:cogasmass}. 
$r_{31}$ is assumed to be 0.53. 
}
\tablenotetext{d}{Tentative detection.}
\tablenotetext{e}{Radio galaxy.}
}
\label{tab:summary}
\end{deluxetable*}
\end{rotatetable*}
%%%%%%%%%%%%%%%%%%%%%%%%%%%%%%%%%%%%%%%%%%%%%%%%%%%%%%%%%%%%%%%%%%%%%%%%%

\section{Result/Discussion}
\label{sec:res}

\subsection{Comparison between CO-based and dust-based molecular gas masses}
\label{subsec:coexcitation}

When using CO(3--2) line to estimate molecular gas masses, 
we need a line ratio between CO(3--2) and CO(1--0) ($r_{31}$). 
The CO excitation state in star-forming galaxies is one of the uncertainties 
in molecular gas mass estimation based on high-{\it J} CO lines. 
Observations of nearby galaxies indicated that the mean line ratio ($r_{31}$) is 
about $0.6$, but it spans a wide range ($0<r_{31}<2$) 
across various galaxies \citep{yao_co_2003}.
\citet{mao_extragalactic_2010} discussed $r_{31}$ by using more than 60 nearby galaxies, 
finding that active galaxies such as starbursts and AGNs tend to have higher $r_{31}$ values 
than normal star-forming galaxies 
\citep{aravena_cold_2010, swinbank_intense_2010, ivison_tracing_2011}.

As for the $r_{31}$ ratio for galaxies at high redshifts, 
some studies reported a similar mean ratio of $r_{31} \sim 0.5$ 
with a large dispersion for a handful of $BzK$ galaxies 
\citep{aravena_co1-0_2014, daddi_co_2015}.
On the other hand, 
\citet{sharon_total_2016} indicated a higher $r_{31}$ value of $r_{31}=0.78\pm 0.27$ for DSFGs at $z\sim 2-3$.
Moreover, \citet{riechers_vla-alma_2020} obtained a mean ratio of $r_{31}=0.91\pm0.14$ 
for the galaxies selected with CO(3-2) line strength, 
suggesting that 
the CO(3-2)-selected galaxies tend to have higher CO line excitation on average.

The CO excitation states of galaxies residing in high-density environments 
are not well-understood yet. 
\citet{coogan_merger_2018} investigated the gas excitation of galaxies 
in the cluster CL J1449+0856 at $z=1.99$.
They found high gas excitation, as seen in the starbursts, 
implying that 
environments can impact the CO excitation states 
through galaxy interactions/mergers. 
%Alternatively, 
%It is also suggested that 
%the compression of gas can cause high gas excitation 
%due to ram-pressure 
a%by intra-cluster matter 
%\citep{bekki_galactic_2013}.

%can be interpreted 
%as the result of compression of gas 
%due to ram-pressure by intra-cluster matter 
%\citep{bekki_galactic_2013}.

\begin{figure}[!tb]
  \centering
  \includegraphics[keepaspectratio, scale=1.0]{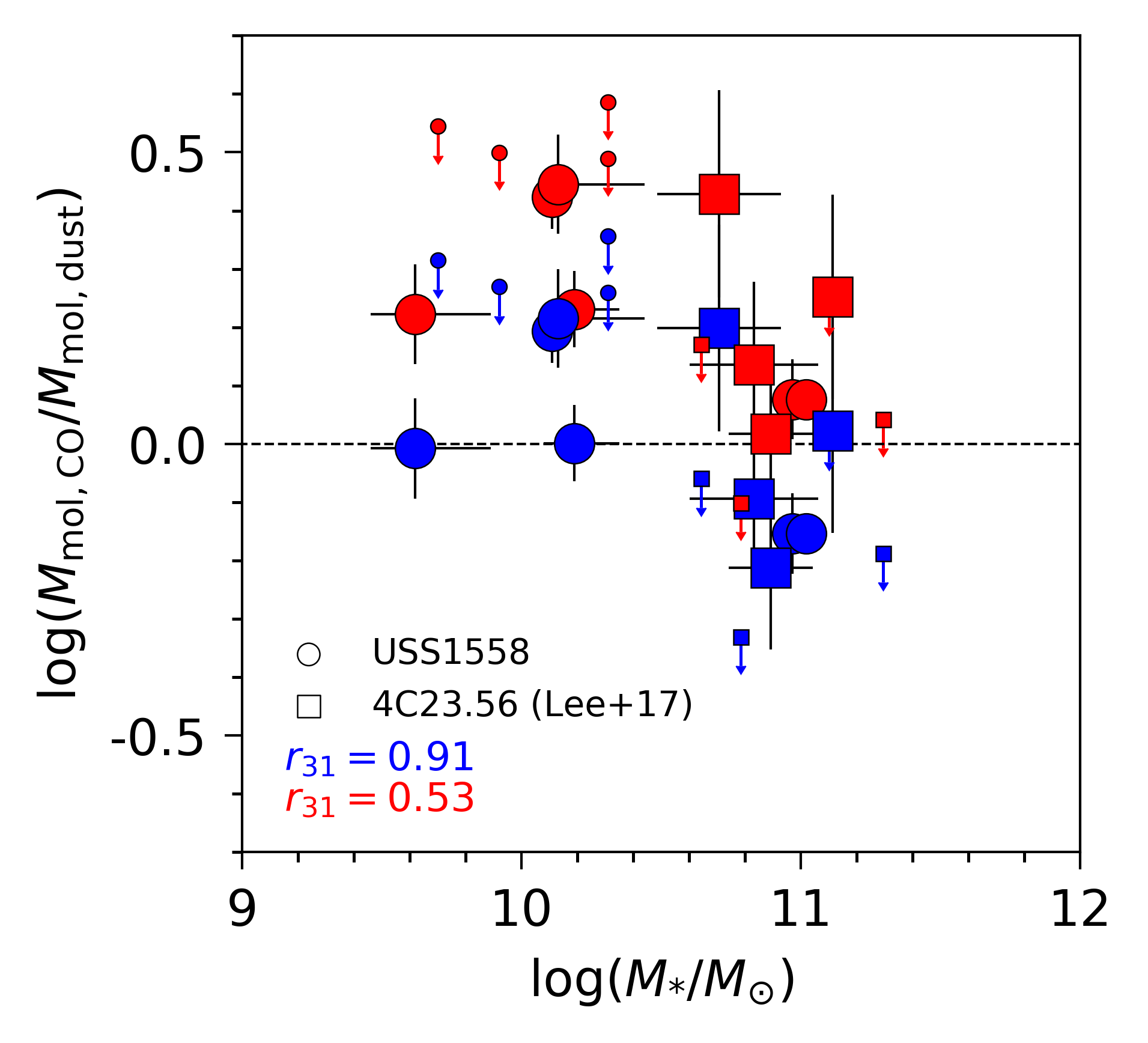}
  \caption{Ratio between the molecular gas masses estimated from 
  CO(3--2) line and dust continuum emission as a function of stellar mass 
  for star-forming galaxies in two protoclusters, USS1558 at $z=2.53$ (circles) and 4C23.56 at $z=2.49$ (squares). 
  %using two different methods are shown by the ratios.
  %Circles and squares show the galaxies in protoclusters, respectively.
  We here assume two different $r_{31}$ values, 
  namely, 0.53 (red symbols; \citealt{lee_radio--mm_2017}) and 
  0.91 (blue symbols; \citealt{riechers_vla-alma_2020}). 
  Assuming $r_{31}=0.53$, the CO(3--2)-based gas mass values are 
  systematically larger than the 
  dust-based gas mass values 
  by $\sim$0.23~dex for the member galaxies in the two protoclusters.
  }
  \label{fig:CO_dust}
\end{figure}

Figure~\ref{fig:CO_dust} shows the difference between the two gas mass measurements 
as a function of stellar mass 
for the HAEs in USS1558. 
We also show star-forming galaxies in a different protocluster at $z\sim2.5$, 4C23.56, from \citet{lee_radio--mm_2017}  
to increase the number of galaxies on this plot. 
In 4C23.56, similar observational data, such as NB imaging data and ALMA Band-3 and 6 data, are available \citep{lee_radio--mm_2017}. 
%for galaxies in two protocluster regions at $z\sim2.5$, 
%namely, USS1558 and 4C23.56 \citep{lee_radio--mm_2017}. 
%\color{red}
%We added the sample 
%in a cluster at a similar redshift 
%from \citet{lee_radio--mm_2017} to enlarge the sample size.
As for the CO(3--2)-based gas masses,
we here assume two different $r_{31}$ values, 
namely, 0.53 \citep{lee_radio--mm_2017} 
and 0.91 \citep[a typical value of CO(3--2)-detected galaxies at $z\sim$ 2--3;][]{riechers_vla-alma_2020}. 
In the case of $r_{31}=0.53$, we find that the CO(3--2)-based gas mass tends to be higher than the dust-based gas mass for the HAEs in the two protoclusters. 
The difference between the two gas measurements appears to become larger for less massive galaxies with $\rm log(M_*/M_\odot) < 10.8$ (Figure~\ref{fig:CO_dust}). 
%%We note that this systematic difference of 0.23~dex cannot be explained by the systematic uncertainty of our gas mass measurement of 0.25~dex (Section~\ref{subsubsec:cogasmass}). 

This result suggests that a higher $r_{31}$ value would be more applicable for the HAEs in USS1558, especially less massive ones, in terms of the consistency between the two gas mass measurements. 
As shown in Figure~\ref{fig:M_SFR}, the HAEs with $\rm log(M_*/M_\odot) < 10.8$ in USS1558 have systematically high SFRs than the MS of star-forming galaxies at the same redshift. 
Given that a positive correlation between the CO excitation and SFR surface density is reported \citep{daddi_co_2015}, the high star-forming activities of the HAEs in USS1558 may lead to a high CO excitation state within them. 
We note that the individual measurements of $r_{31}$ are necessary in order to confirm whether the HAEs in USS1558 have a high CO excitation state on average and to investigate how the excitation states correlate with their star-forming activities. 

\citet{tadaki_environmental_2019} estimated the CO(3--2)-based gas masses of the HAEs in USS1558 assuming $r_{31} = 0.56$. 
When a higher $r_{31}$ is more applicable for the HAEs in this protocluster as suggested above, their gas mass values might be overestimated by a factor of 1.5.

\subsection{Comparison with field galaxies on the MS}
\label{subsec:scale_MS}

\begin{figure*}[!tbp]
  \centering
  \includegraphics[width=0.9\textwidth]{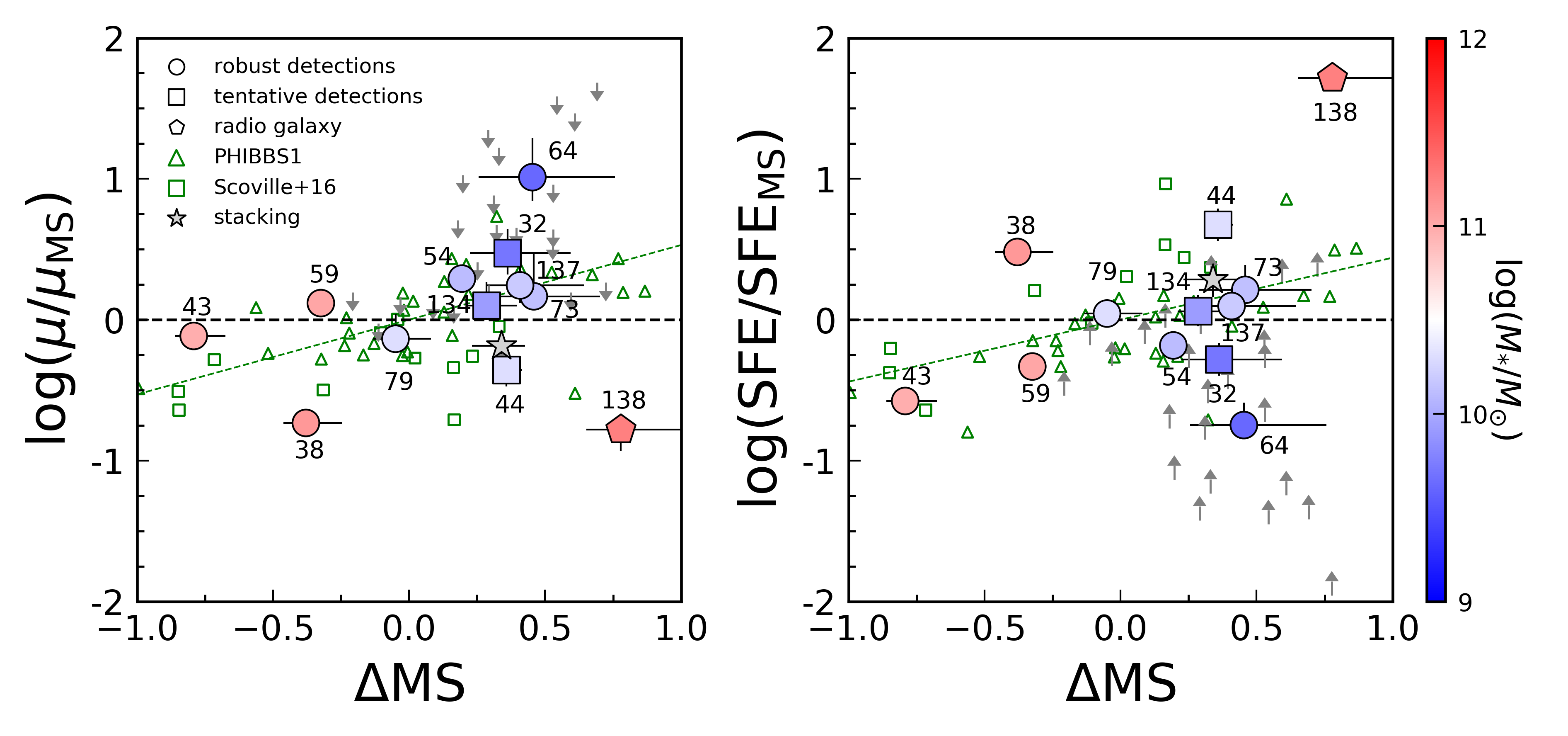}
  \caption{
  Normalized dust-based gas mass ratio (left) and SFE (right) 
  as a function of the deviation from the star-forming main sequence, $\Delta \rm MS = \log(\rm sSFR/sSFR_{MS})$.
  %The meanings of different symbols are given in the legend, and the color of the symbols
  %corresponds to stellar mass as shown in the color-scale bar on the right side.
  Gas mass ratio and SFE are normalized 
  by the expected values from the gas scaling relation 
  \citep{tacconi_phibss_2018} at $\Delta \rm MS=0$.
  %This is equivalent to a comparison of 
  %the protocluster galaxies with the general field galaxies 
  %on the main sequence with the same stellar mass. 
  The HAEs in USS1558 are shown with large circle and square symbols and color-coded according to their stellar masses. 
  Green open triangles represent the field galaxies 
  at $z=2$--2.5 from the PHIBBS1 survey \citep{tacconi_phibss_2013} 
  and green open squares represent the field galaxies 
  at $z=1.5$--3.0 from \citetalias{scoville_ism_2016}.
  The green dashed lines show the scaling relations for the general field galaxies \citep{tacconi_phibss_2018}.
  Among the HAEs in USS1558, we find a trend that the molecular gas mass ratio and SFE both increase with increasing $\Delta \rm MS$, which suggests that the large gas reservoir within the galaxies sustains the elevated star-forming activities observed in the dense cores of USS1558. 
  Comparing with the scaling relations, our protocluster galaxies appear to form stars according to their gas contents just like the filed galaxies equally above the MS do. 
  %Our protocluster galaxies tend to be located above the main sequence ($\Delta \rm MS>0$) due to their higher star formation activities, but they still seem to follow the same relations for the general field galaxies where the molecular gas mass ratio and the star formation efficiency both increase with increasing $\Delta \rm MS$.
  %The protocluster galaxies tend to be slightly more gas rich than the field MS galaxies,
  Note that HAE 138 is the radio galaxy and neither its SFR nor gas mass can be accurately measured due to the contamination of luminous AGN in the observed fluxes. Therefore, this data point is discounted.
}
  \label{fig:distri}
\end{figure*}

Figure~\ref{fig:distri} shows the gas properties, namely, 
gas mass to stellar mass ratio ($\mu=M_{\rm mol}/M_*$) and star-formation efficiency (SFE$={\rm SFR}/M_{\rm mol}$), 
as a function of 
the deviation from the MS ($\Delta \rm MS$).
Those quantities are normalized by the expected values from the gas scaling relation 
\citep{tacconi_phibss_2018} 
with $\Delta \rm MS=0$. 
This is equivalent to a comparison of 
protocluster galaxies to the field galaxies 
on the MS at a given stellar mass.
We find that the protocluster galaxies with larger $\Delta \rm MS$ tend to be more gas-rich and have higher SFEs, but 
 they still seem to follow the same relations for the field galaxies shown with the dashed line in Figure~\ref{fig:distri}. 
%The protocluster galaxies (HAEs in USS1558) tend to be located above the main sequence ($\Delta \rm MS>0$) due to their higher star formation activities \citep{shimakawa_mahalo_2018}, but they still seem to follow the same relations for the general field galaxies where the molecular gas mass ratio and the star formation efficiency both increase with increasing $\Delta \rm MS$.
The protocluster galaxies tend to be slightly more gas-rich than the field MS galaxies, 
but they form stars according to their gas contents 
just like the field galaxies
equally above the main sequence do.

The slightly higher gas mass ratios seen for protocluster galaxies might be partly attributed 
to the effective cold accretion by cold streams 
along the surrounding filamentary structures.
At the earlier stage of cluster formation, 
cold gas can be efficiently supplied through cold streams 
penetrating the hot ICM \citep{dekel_cold_2009}, 
while such cold mode accretion is prevented 
by hot ICM in matured clusters such as X-ray clusters.
The dynamical masses of F1 and F2 groups are estimated 
to be $0.10\times10^{14}M_{\odot}$ and $0.87\times10^{14}M_{\odot}$, respectively, 
assuming a local virialization 
\citep{shimakawa_identification_2014}.
This protocluster is thought to be 
yet in the accretion-dominated phase
before it becomes to an accretion-inefficient phase 
\citep{dekel_cold_2009}.
In addition, \citet{shimakawa_direct_2017} found that
Ly$\alpha$ emitters in this field are distributed so as to avoid 
the densest protocluster core regions traced by HAEs. 
This indicates that 
Ly$\alpha$ emissions from those HAEs in the densest regions 
are suppressed by the associated, abundant {\sc Hi} gas 
and/or larger dust contents in the HAEs 
in the densest regions. 
This result also supports the idea that 
this protocluster is in a young, accretion-dominated phase, 
and thus, member galaxies can efficiently obtain gas from the outside.

Galaxy interactions or mergers are also expected to be happening 
in the cores of this protocluster.
In dense regions, a merger rate is naturally expected 
to be higher than that in the field. 
\citet{jian_environmental_2012} reported such a trend 
using semi-analytic models and showed that 
the merger rate peaks in the halos 
in the mass range of $10^{12-13}h^{-1}M_{\odot}$, 
which corresponds to group environments.
In fact, HAE32 and HAE64, 
which are the most gas-rich galaxies among our sample 
(Table~\ref{tab:summary} and Figure~\ref{fig:distri}), 
are accompanied by another HAE. 
The projected distance to the companion is $0''.42$ and $1''.47$ 
for HAE32 and HAE64, respectively, 
corresponding to 3.4 and 11.9~kpc at $z=2.53$, 
and therefore the companions are likely to be interacting 
with the host HAEs.
Moreover, \citet{tadaki_evidence_2014} pointed out the possibility that 
HAE54 and HAE59 are interacting with each other
based on the CO(1-0) observations with Jansky Very Large Array (JVLA).
These interacting galaxies tend to show higher gas mass ratios 
compared to normal star-forming galaxies on the MS 
(the left panel of Figure~\ref{fig:distri}).

A recent study by \citet{moreno_spatially_2020} investigated 
the evolution of star formation and ISM in interacting galaxies 
using FIRE-2 simulations. 
They showed that the molecular gas budget increases 
during the galaxy-pair or the interaction-phase before coalescence. 
\citet{violino_galaxy_2018} also reported the molecular gas enhancement ($\sim 0.4$ dex) 
in pair galaxies at $z\sim 0.03$ 
selected from the Sloan Digital Sky Survey Data Release 7 
\citep[SDSS DR7,][]{abazajian_seventh_2009}.
One of the plausible mechanisms 
to explain enhanced molecular gas during the interaction-phase is that 
the gas is driven into the galaxy centers by losing the angular momentum 
and the dense molecular gas is formed 
%interactions drive the gas that has lost the angular momentum 
%into the galaxy centers 
%and concentrated HI gas is converted into molecular gas 
\citep{braine_co1-0_1993, combes_co_1994, moreno_mapping_2015}.
Based on these studies, 
we suggest that the protocluster galaxies which have close companions are 
at the early stage of interactions and their high gas mass ratios
are driven by the interactions.
%ZZZ what is the mechanism? I guess interactions drive the gas that has lost the angular momentum into the galaxy centers and form dense molecular cores. Better to write it here with a reference. ZZZ

For massive galaxies with $\log(M_{*}/M_{\odot})>11$, 
except for the RG, 
their star formation rates are below the MS (Figure~\ref{fig:M_SFR}).
One possibility to explain this trend is that  
the assumption of 
$E(B-V)_{\rm nebular} = E(B-V)_{\rm stellar}$
when estimating H$\alpha$-based SFRs (Section~\ref{subsubsec:SFR measurement}) 
is inappropriate for the massive HAEs.
The ratio of $E(B-V)_{\rm nebular}/E(B-V)_{\rm stellar}$ is suggested to become larger for more massive galaxies in the local universe \citep{koyama_different_2019}.
When this is also the case for the massive HAEs in USS1558, their dust extinction values for H$\alpha$ 
would be underestimated. 
Another possibility is that 
their low SFRs relative to the MS 
are intrinsic signatures and that 
the massive HAEs are in the transitional phase 
to quiescent galaxies. 
As discussed in \citet{tadaki_environmental_2019}, 
given that the massive galaxies reside in massive dark matter halos, 
the cold gas accretion to massive galaxies would be suppressed 
due to the virial shock heating, 
which would then lead to the suppression of their star-forming activities. 

Based on the stacking analyses of non-detected galaxies 
with $9.5< \log(M_{*}/M_{\odot})<10.0$, 
their averaged gas mass ratio is 
relatively low 
while the averaged SFE is comparable to that in the field.
A part of the molecular gas in these less massive galaxies might be removed via galaxy harassment, halo gas stripping, or ram pressure stripping, although it is still under debate whether the ram pressure stripping could affect molecular gas components in galaxies. 
%%
%A part of the molecular gas in these less massive galaxies 
%might be stripped by interactions with massive galaxies, 
%\color{red}
%although it is still under debate whether the stripping process could affect molecular gas components in galaxies 
%\color{black}
%although it is still under debate whether 
%such a process is realistic or not 
%\citep{kenney_effects_1989, boselli_cold_2014}.
%
The relatively low gas mass ratio of the stacked sample 
might be driven by the fact that 
the sample includes less active, and thus, 
gas-poorer HAEs on and below the MS  (Figure~\ref{fig:M_SFR}). 
Deeper observations and thus individual gas mass measurements  
are needed for further discussion 
on the gas properties of those less massive galaxies.

\subsection{Comparison with gas scaling relation}

\label{subsec:scaling}
\begin{figure*}[!bt]
  \centering
  \includegraphics[width=\textwidth]{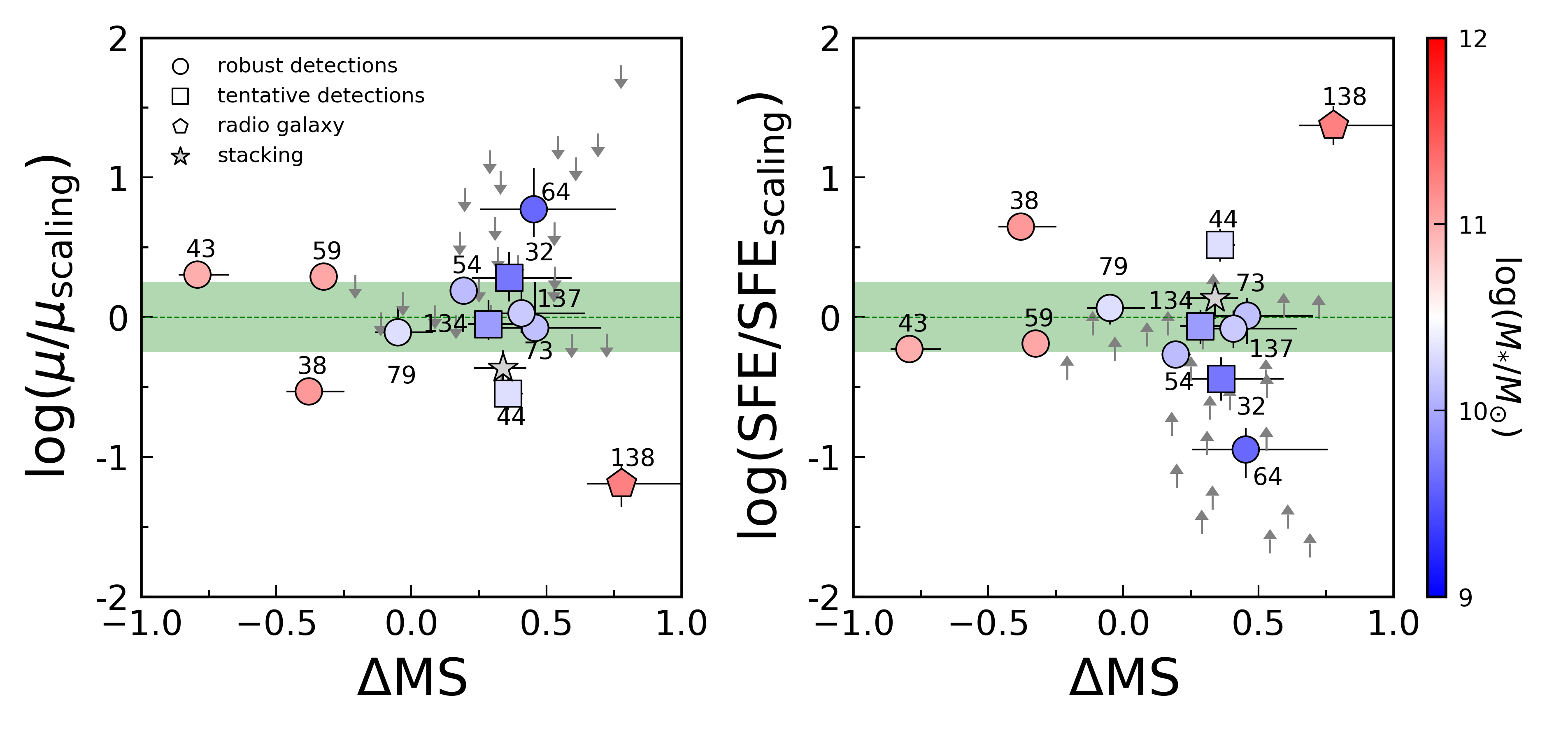}
  \caption{Normalized dust-based gas mass ratio (left) and SFE (right) 
  as a function of stellar mass for the HAEs in USS1558. 
  %The meanings of different symbols are the same as in Fig.~\ref{fig:distri}.
  The symbols are color-coded according to their stellar masses. 
  Gas mass ratio and SFE are normalized 
  by the values inferred from 
  the gas scaling relation \citep{tacconi_phibss_2018} 
  at the same stellar mass and SFR (offset from the MS).
  %This is equivalent to a comparison of the protocluster galaxies with the general field galaxies with the same stellar mass and SFR. 
  The green shaded region corresponds to 
  a scatter of the field scaling relation \citep{tacconi_phibss_2018}. 
  Although the scatter of our sample is large, 
  many of the protocluster galaxies appear to have 
  similar gas mass ratios and SFEs to 
  those of field galaxies with the same stellar mass and SFR. 
  }
  \label{fig:scaling}
\end{figure*}

We investigate 
the environmental dependence of gas properties 
by comparing with the field galaxies' scaling relation 
at a fixed stellar mass and SFR (offset from the MS). 
Figure \ref{fig:scaling} shows 
the gas mass ratio and SFE %(inverse of gas depletion timescale) 
as a function of stellar mass. 
Gas properties are normalized by the values inferred from the scaling relation of general field galaxies \citep{tacconi_phibss_2018} 
at the same stellar mass and the same SFR (or $\Delta \rm MS$). 
Many of the protocluster galaxies have gas mass ratios 
comparable to those in the field 
within the intrinsic scatter of the field scaling relation of $\pm 0.25$ dex. 
Similarly, even though there are some galaxies 
that are significantly offset from the scaling relation, 
our protocluster galaxies tend to have similar SFEs 
as field galaxies.
%as a whole. ZZZ we may not say "as a whole"?
These results suggest that 
there is no significant environmental dependence 
in gas properties 
if we compare galaxies at the same stellar masses and SFR. 
In other words, 
the relation 
between star formation activity and gas content 
can be universal across the environment at $z\sim2.5$.

\subsection{Comparison with other protoclusters}
Recently, there is an increasing number of studies
on gas properties of distant clusters in the literature.
However, how the environments can actually affect or 
alter the gas properties of galaxies 
at $z\sim$ 2--3 is still highly controversial \citep{tacconi_evolution_2020}.  
\citet{wang_revealing_2018} investigated 
the gas properties of galaxies 
in the most distant X-ray cluster, CLJ1001 at $z=2.51$, 
by conducting CO(1--0) line observations with JVLA.
They found a strong environmental dependence of gas properties 
on the clustercentric radius. 
They discussed the environmental effects such as 
stopping gas accretion onto galaxies or 
removing gas out of galaxies 
due to ram-pressure or tidal force.
Another X-ray cluster, J1449+0856 at $z$=1.99, was also observed by \citet{coogan_merger_2018} with ALMA and JVLA, and it showed a deficit of gas mass fractions in member galaxies.
\citet{tadaki_environmental_2019} showed 
an enhancement of gas mass in less-massive galaxies 
in three protoclusters at $z\sim$ 2--2.5, including USS1558.
They discussed that higher gas fraction 
in the protocluster galaxies
is due to efficient cold gas accretion via cosmic filaments.
\citet{gomez-guijarro_2019} also showed two Herschel selected protoclusters at $2<z<2.6$ and showed the enhancement of gas mass fractions based on ALMA CO(3--2) line observations, which lead to the enhancement in star formation rate above the main-sequence, while keeping the star formation efficiency almost constant.
This result is fully consistent with ours.
On the other hand, 
\citet{lee_radio--mm_2017} found 
no clear environmental dependence 
of the gas properties of seven CO(3--2)-detected galaxies 
in 4C23.56 protocluster at $z=2.49$. 
Also, \citet{zavala_gas_2019} investigated 
two protoclusters at $z=2.10$ and 2.49
in the COSMOS field and showed that 
the gas properties of the protocluster galaxies are consistent 
with the field scaling relation.
Moreover, \citet{darvish_similar_2018} found that 
the gas properties do not depend on 
the local density using the sample of \citet{scoville_evolution_2017}. 
%These contradicting results mean that 
%the environmental dependence of gas properties 
%in different clusters 
%has not reached any consensus yet.

The contradicting results mentioned above may reflect 
the difference in the evolutionary stage of (proto)clusters, 
considering that the dominant physical processes 
working on member galaxies are expected to vary 
depending on the evolutionary stages of (proto)clusters 
%in different clusters with various properties 
%such as mass and age 
\citep{shimakawa_mahalo_2018-1}.
%The mode of gas accretion 
%are related to these properties and 
%have a vital impact on gas properties of (proto)cluster galaxies 
%\citep{dekel_galaxy_2006}.
For instance, in matured clusters with extended X-ray emission, 
the gas within cluster galaxies can be stripped by ram-pressure and turn out to be gas-poor.
%resulting in suppressing or truncating star formation.
In fact we see a deficit in gas fractions for the above two X-ray clusters (CLJ1001 and J1449).
On the contrary, in immatured (proto)clusters, (proto)cluster galaxies can acquire fresh gas through cold accretion and be more gas-rich with respect to the field galaxies. 
%possibly leading to vigorous star formation \citep{dekel_galaxy_2006}.
%Different mechanisms at different accretion phase 
%working on galaxy clusters 
%produce the opposite results for gas properties.
%This is consistent with 
%the observed cluster-to-cluster large variations 
%found at $z>2$ \citep{tacconi_evolution_2020}.

Our USS-1558 protocluster is considered to be in a growing phase, 
where fragmented groups are hierarchically assembling 
and merging to form a single viriarized cluster.
We suggest that, at this early assembly stage, 
cold accretion is vigorously occurring, and 
the accreted gas is converted to stars to sustain 
their high star formation activities, 
but in a way that the relation 
between molecular gas properties and star formation activities 
is kept the same following the scaling relation.

\subsection{Relation between galaxy size and gas properties}

\begin{figure*}[!htbp]
  \centering
  \includegraphics[width=\textwidth]{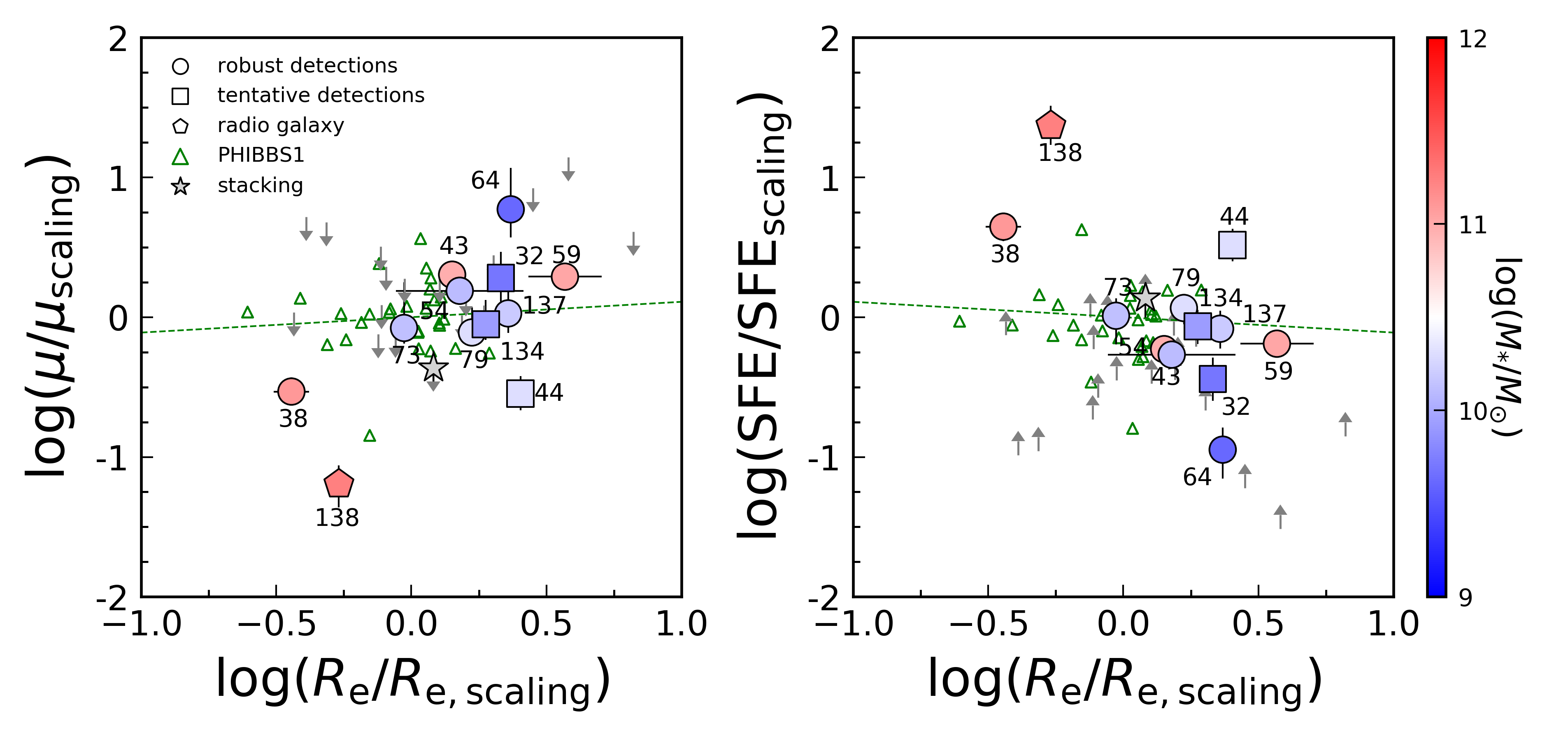}
  \caption{
  Normalized dust-based gas mass ratio (left) and SFE (right) 
  as a function of half-light radius 
  normalized by the value 
  from the mass-size relation of
  \citep{van_der_wel_3d-hstcandels_2014} at a fixed stellar mass. 
  Gas mass ratio and SFE are normalized by the  values inferred from the gas scaling relation \citep{tacconi_phibss_2018} 
  at a fixed stellar mass and SFR. 
  The HAEs in USS1558 are color-coded according to their stellar masses. 
  Green open squares represent the filed galaxies from the PHIBSS1 survey \citep{tacconi_phibss_2013}. 
  A green dashed lines represent the size dependence of gas properties 
  found in \citet{tacconi_phibss_2018}. 
  Our protocluster galaxies show a similar dependence of 
  gas properties on size as field galaxies 
  although there are some outliers. 
  }
  \label{fig:size}
\end{figure*}

The galaxy size is an important property for galaxies' characterization and further understanding of environmental effects.
\citet{tacconi_phibss_2018} found a weak size dependence of 
gas properties in field galaxies.
The size dependence is almost negligible considering that 
the slope of 
$d\log(\mu/\mu_{\rm scaling})/d\log(R_{\rm e}/R_{\rm e,\,scaling}) = 0.11$, $d\log(\rm SFE/SFE_{scaling})/d\log(R_{\rm e}/R_{\rm e,\,scaling}) = -0.11$ 
and the intrinsic scatter of mass-size relation is 
$\gtrsim0.2$ for all galaxy types and redshifts 
\citep{van_der_wel_3d-hstcandels_2014}. 

Here, we investigate 
the size dependence of gas properties 
for the protocluster galaxies.
Figure~\ref{fig:size} shows the relation 
between size and gas properties.
Whereas the overall trend of the protocluster galaxies follows 
the trend seen in the field galaxies, 
the protocluster galaxies have a large scatter
compared to the field galaxies from PHIBBS1 
\citep{tacconi_phibss_2013}.   
We find some outliers among the protocluster galaxies. 
This suggests that these outliers experience peculiar physical processes such as galaxy mergers.

In particular, HAE38 has a remarkably compact size 
($R_{\rm e}=1.47$~kpc)
and shows a relatively low gas mass ratio ($\rm \mu_{mol}=0.16$) and high SFE ($\rm log(SFE\, [Gyr^{-1}])=$0.7) among the HAEs in USS1558.
%Its gas properties are likely to be those of starburst galaxies in spite of its less-active star formation. 
\citet{spilker_2016_low} investigated the gas reservoir in massive and compact star-forming galaxies at $z\sim2.3$ in the CANDELS fields with CO(1--0) emission line. 
The gas mass ratios and SFE of the compact star-forming galaxies are $\mu_{\rm mol}\lesssim0.14$ and ${\rm log(sSFR\, [Gyr^{-1}])}\gtrsim1.0$, respectively (including one CO-detected source).
The gas properties as well as size of HAE38 appear to be similar as those of the massive and compact star-forming galaxies at similar redshifts \citep{spilker_2016_low}. 
Given the relatively low active star-formation of HAE38 (Figure~\ref{fig:M_SFR}), 
%starburst-like gas properties in spite of its less active star formation.
HAE38 might be in a transition phase 
from compact and massive star-forming galaxies 
to compact and massive quiescent galaxies, 
the so-called ``red nugget'' \citep[e.g.,][]{daddi_passively_2005,damjanov_red_2009}. 
Compact and massive star-forming galaxies 
are suggested to be formed by gas-rich dissipational processes, such as gas-rich galaxy-galaxy mergers or violent disk instabilities caused by intense gas inflow 
(e.g., \citealt{barro_candels_2013,dekel_wet_2014}, but see also \citealt{van_dokkum_forming_2015}). 
The starburst-like gas properties and the compact size of HAE38 
are likely to be consistent with this scenario.

RG (HAE138) has a small radius probably because the emission from the central bright AGN is dominant.
Galaxies in the interaction-phase such as HAE32 and HAE64 have relatively large radii compared to expected values 
from the mass-size relation of 
\citet{van_der_wel_3d-hstcandels_2014}.
This indicates that these interacting galaxies are physically disturbed 
and thus have extended structures or central regions of these galaxies are highly affected by dust attenuation and show higher effective radii than intrinsic ones. 

\section{Summary}
\label{sec:summary}

In this study, we conducted the ALMA Band-6 observations 
of the dense cores of the protocluster USS1558 at $z=2.53$,  
to investigate the environmental dependence of gas properties, and to understand the origin of the star formation enhancement 
seen in those galaxies.
We detected interstellar dust emissions from 12 member galaxies and 
estimated their molecular gas masses 
based on the Rayleigh-Jeans dust continuum fluxes 
\citepalias{scoville_ism_2016}.
We summarize our findings as follows:

\begin{itemize}
 
\item 
We compared the dust-based molecular gas mass and 
the CO-based molecular gas mass 
to check the consistency
between the two independent gas mass measurements.
We found that the CO(3--2)-based gas masses are 
systematically offset upward ($\sim0.2$ dex) 
from the dust-based gas masses 
when the CO line ratio of $r_{31}=0.53$ is assumed. 
This offset is reduced if a higher line ratio is assumed, 
which suggests that the molecular gas in protocluster galaxies 
may be more excited due to 
high star formation activities.
%the environmental effects like galaxy-galaxy %%interactions/mergers \citep{coogan_merger_2018} or 
%ram-pressure from ICM \citep{bekki_galactic_2013}.

\item 
When the HAEs in USS1558 have a higher CO gas excitation state as suggested from the comparison of the two gas mass measurements, our previous work \citep{tadaki_environmental_2019}, in which $r_{31}$=0.53 is assumed to estimate the molecular gas masses from CO(3--2) line, might overestimate the gas masses by a factor of 1.5.

\item 
The enhancement of both star formation and gas content 
can be seen in star-forming member galaxies with $\log(M_*/M_\odot)>10.0$ at least.
This suggests that the high gas mass fraction led by cold accretion 
along the filamentary structures is sustaining 
their enhanced star-forming activities.
Besides, galaxy interactions are also expected to be occurring 
more frequently in the protocluster cores 
due to the high number density of galaxies, 
and contributing to the elevated star formation activities 
to some extent.

\item 
There is no significant difference between 
protocluster galaxies and coeval field galaxies 
if we compare gas properties at fixed $M_{*}$ and SFR 
(or $\Delta \rm MS$).
This result indicates that the relationship between 
gas content and star formation activity is 
universal even in the densest protocluster cores.

\item
We investigated the dependence of gas properties 
on galaxy sizes.
While a clear correlation is not seen in the protocluster galaxies 
as well as in the coeval field galaxies, 
there are some outliers that are offset from normal size, 
suggesting that 
some galaxies are evolving via the peculier processes such as mergers.

\end{itemize}

Our deep observations with ALMA revealed the global gas properties 
in protocluster galaxies with $\log(M_{*}/M_{\odot})>10$.
Because the number of individual detections of dust continuum is not 
yet sufficient for a statistical discussion, 
more observations of protoclusters at high redshifts are needed 
to obtain a global picture of the environmental impacts 
on the gas properties of galaxies in the early Universe. 
Additionally, CO(1-0) observations are required 
to reveal the gas excitation and kinematics.
Moreover, gas kinematics observation would be ideal as 
it will give us valuable information on the physical processes 
at work in galaxies in young protoclusters. 
The information of atomic gas ({\sc Hi}) associated to this field is 
also important to explore 
the mode of gas accretion of this protocluster.
{\sc Hi} tomographic mapping will enable us to 
reveal the large-scale structure of {\sc Hi} gas associated with 
the protocluster cores and to discuss the relation between 
the surrounding {\sc Hi} gas structures and the gas contents 
in individual member galaxies. 

\acknowledgments
This work was supported by NAOJ ALMA Scientific Research Grant Numbers 2018-08A.
This paper makes use of the following ALMA data: ADS/JAO.ALMA\#2015.1.00395.S, ADS/JAO.ALMA\#2016.1.00461.S, ADS/JAO.ALMA\#2017.1.00506.S. 
ALMA is a partnership of ESO (representing its member states), NSF (USA) and NINS (Japan), 
together with NRC (Canada), MOST and ASIAA (Taiwan), and KASI (Republic of Korea), in 
cooperation with the Republic of Chile. The Joint ALMA Observatory is operated by 
ESO, AUI/NRAO and NAOJ.
Data analyses were in part carried out on the open use data
analysis computer system at the Astronomy Data Center, ADC, of 
the National Astronomical Observatory of Japan.
TK acknowledges the support by Grant-in-Aid for Scientific Research (A) (KAKENHI \#18H03717).

\vspace{5mm}
\facilities{ALMA}
\software{{\sc casa} \citep{mcmullin_casa_2007}
          }
          
\vspace{100mm}

\bibliography{reference2}{}
\bibliographystyle{aasjournal}

\end{document}